\documentclass{aastex}
\usepackage{emulateapj5}
\usepackage{longtable}
\usepackage{amssymb}

\newif\ifAMStwofonts
\AMStwofontstrue

\def\kms{km~s$^{-1}$}

\def\ga{\mathrel{\hbox{\rlap{\hbox{\lower4pt\hbox{$\sim$}}}\hbox{$>$}}}}
\def\la{\mathrel{\hbox{\rlap{\hbox{\lower4pt\hbox{$\sim$}}}\hbox{$<$}}}}

\shorttitle{Atomic gas around Perseus}

\shortauthors{S.\ Stanimirovi\'{c} et al.}
\begin{document}

\title{Cold and warm atomic gas around the Perseus molecular cloud I: Basic Properties}

\author{Sne\v{z}ana Stanimirovi\'{c}\altaffilmark{1}, Claire Murray\altaffilmark{1},
 Min-Young Lee\altaffilmark{2}, Carl Heiles\altaffilmark{3}, 
Jesse Miller\altaffilmark{4,1}}
\altaffiltext{1}{Department of Astronomy, University of Wisconsin, Madison, WI
  53706; sstanimi\@astro.wisc.edu}
\altaffiltext{2}{Laboratoire AIM, CEA/IRFU/Service d'Astrophysique, Bat 709, 91191 Gif-sur-Yvette, France}
\altaffiltext{3}{Department of Astronomy, UC Berkeley, 601 Campbell Hall,
Berkeley, CA 94720}
\altaffiltext{4}{Department of Physics and Astronomy, Washington State University, PO Box 642814, Pullman WA 99164-2814}

\slugcomment{Accepted for publication in ApJ}

\begin{abstract}
Using the Arecibo Observatory we have obtained neutral hydrogen (HI) absorption and emission spectral pairs 
in the direction of 26 background radio continuum sources in the vicinity of the Perseus molecular
cloud. Strong absorption lines were detected in all cases allowing us to 
estimate spin temperature ($T_s$) and optical depth for 107 individual Gaussian components
along these lines of sight. 
Basic properties of individual HI clouds (spin temperature, optical depth, and the column density
of the cold and warm neutral medium, CNM and WNM) in and around Perseus are very similar
to those found for random interstellar lines of sight sampled by the Millennium HI survey.
This suggests that the neutral gas found in and around molecular clouds is not atypical. 
However, lines of sight in the vicinity of Perseus have on average a higher total HI column density
and the CNM fraction, suggesting an enhanced amount of 
cold HI relative to an average interstellar field.
Our estimated optical depth and spin temperature are in stark contrast with the recent attempt at
using Planck data to estimate properties of the optically thick HI.
Only $\sim15$\% of lines of sight in our study have a column density weighted average
spin temperature lower than 50 K, in comparison with $\ga85$\% of Planck's sky coverage.
The observed CNM fraction is inversely proportional to the optical-depth weighted
average spin temperature, in excellent agreement with the recent numerical
simulations by Kim et al.
While the CNM fraction is on average higher around Perseus relative to a 
random interstellar field, it is generally low, $10-50$\%. 
This suggests that extended WNM envelopes
around molecular clouds, and/or significant mixing of CNM 
and WNM throughout molecular clouds, 
are present and should be considered in the models of molecule and star formation.
Our detailed comparison of  HI absorption with CO emission spectra
shows that only 3/26 directions are clear candidates for probing 
the CO-dark gas as they have $N(HI)>10^{21}$ cm$^{-2}$ yet no detectable CO emission.
\end{abstract}


\section{Introduction}

Most of molecular gas in galaxies is assembled into giant molecular clouds (GMCs)
with masses of 10$^{4}$--10$^{7}$ M$_{\odot}$ \citep{Fukui10}.
Stars appear intimately associated with the dense regions of these GMCs \citep{Lada10},
and recent observations suggest
that the depletion timescale of molecular gas by star formation
does not vary greatly across a wide range of galaxy
environments \citep{Schruba11,Shetty14}.
This strongly suggests that the ability to form molecular gas in the first place
holds the key to understanding the evolutionary tracks of galaxies.

Atomic hydrogen has been considered for decades as the main formation reservoir of GMCs
\citep{Shu73, Blitz07, Kim06, Audit05, Heitsch05, Clark12}.
Although how exactly GMCs form out of the diffuse atomic medium is still not understood,
the HI envelopes frequently observed around GMCs are likely to represent the material 
left over from the formation epoch and/or a product of 
photodissociation of molecular gas. In either case,
these envelopes  play a very important role in the GMC evolution and
could explain long-standing questions such as the origin 
of the internal turbulent energy in GMCs. 
Theoretical models considering the ongoing
accretion of atomic material from the envelope onto GMCs 
are able to reproduce the level of observed turbulence,
as well as the total GMC mass \citep{Chieze89, Hennebelle06, Goldbaum11}.
In addition, it has been suggested that the GMC history is highly dependent on the 
initial surface density of the HI envelope. As shown by Goldbaum et al. (2011), 
only a factor of two increase of the  HI surface density of the 
envelope from 8 to 16 M$_{\odot}$ pc$^{-2}$
is enough to decide whether or not a GMC mass will reach  $\sim10^6$ 
M$_{\odot}$ over a typical lifetime of 10-20 Myr.

While the HI envelopes around molecular clouds have been 
largely observationally studied via HI emission 
\citep{Wannier83, Wannier91, Andersson93, Fukui09}, 
traditionally HI has not been considered as 
very important for understanding molecule and star formation. For example, many GMC studies 
trying to estimate the H$_2$ distribution from dust emission have neglected 
to account for HI as it was assumed that GMCs are highly dominated by
molecular gas (e.g. \cite{Pineda08}).  In addition, 
a strong correlation between the star formation rate
and the H$_2$ surface density in galaxies has been considered as an evidence that 
only H$_2$ is directly related to star formation.
However, recent 
extragalactic studies showed that globally across galaxies at kpc-scales, as well as
in resolved studies at sub-kpc scales, the HI surface density
$\Sigma_{\rm H\textsc{i}} \la $10 M$_{\odot}$ pc$^{-2}$
\citep{Wong02, Blitz04, Bigiel08,Schruba11},
re-opening interest in the role of HI shielding in molecule formation.

To investigate the formation of H$_2$
from a theoretical point of view,
and building up on several earlier studies \citep{Spitzer75,Elmegreen93},
\cite{Krumholz09} (KMT09) considered the structure of a photodissociation region (PDR) 
in a spherical cloud that is embedded in a uniform and isotropic radiation field. 
Their model is based on the balance between H$_{2}$ formation on dust grains 
and photodissociation by Lyman--Werner (LW) photons 
and provides an analytic function for the H$_{2}$ fraction as a 
function of the gas surface density. 
Their most important prediction is that a certain amount of the H\textsc{i} surface density, 
$\Sigma_{\rm H \textsc{i}}$, is required for shielding of H$_{2}$ against photodissociation. 
Once this minimum $\Sigma_{\rm H\textsc{i}}$ is achieved, 
additional H\textsc{i} is fully converted into H$_{2}$ and 
therefore $\Sigma_{\rm H\textsc{i}}$ saturates
while $\Sigma_{\rm H2}$ linearly increases. 
At solar metallicity, KMT09 predict $\Sigma_{\rm H\textsc{i}} \sim 10$ M$_{\odot}$ pc$^{-2}$ 
as the minimum $\Sigma_{\rm H\textsc{i}}$ required for H$_{2}$ formation,
this is equivalent to the HI column density of $1.2\times10^{21}$ cm$^{-2}$.

To investigate the role of HI shielding on sub-pc scales in \cite{Lee12} 
we mapped the transition from H\textsc{i} to H$_{2}$ across the Perseus
molecular cloud\footnote{Perseus is located at a distance of 200--350 pc 
\citep{Herbig83}, has
$M \sim 2 \times 10^{4}$ M$_{\odot}$ \citep{Sancisi74,Lada10} and solar metallicity
\citep{Gonzalez09}.}.
The HI data in this study were from the GALFA--H\textsc{i} survey 
\citep{Peek11, Stanimirovic06}
and the HI column density was estimated under the optically thin assumption.
To estimate the H$_2$ image, the 60 and 100 $\mu$m data from the Improved Reprocessing
of the \textit{IRAS} Survey (IRIS) \citep{Miville-Deschenes05} were used.
We derived 
$T_{\rm dust}$ from the $I_{60}/I_{100}$ ratio, and
then converted $\tau_{100}$ to $A_{V}$ by finding 
a proportionality constant between our derived $A_{V}$ and the 
$A_{V}$ image derived from optical extinction (provided by the 
COMPLETE survey, \cite{Ridge06}).
Finally, the H$_{2}$ column density was calculated as: $N$(H$_{2}$)=
($A_{V}$/DGR $-$ $N$(H\textsc{i}))/2;
the dust-to-gas ratio DGR = 1.1 $\times$ 10$^{-21}$ mag cm$^{2}$ was measured locally around  Perseus.

The key result from Lee et al. (2012) is
the detection of an almost constant $\Sigma_{\rm H\textsc{i}}$ of 6--8 M$_{\odot}$ pc$^{-2}$ 
for several dark and star-forming regions in Perseus.
This is in agreement with KMT09's prediction for the saturation of $\Sigma_{\rm H\textsc{i}}$.
In addition, Lee et al. showed that H$_2$ extends up to 20 pc from core centers,
and that the HI envelope is very extended ($>20$ pc). 
The HI halo of Perseus was previously studied by \cite{Andersson93}
who focused on dark region B5.  Using radiative transfer modeling they found that 
the HI halo is about $5\times8$ pc in size.

While the observed flattening of $\Sigma_{\rm H\textsc{i}}$ can be
attributed to the conversion of H\textsc{i} into H$_{2}$ as in KMT09,
an alternative possibility is that $\Sigma_{\rm H\textsc{i}}$ is
simply underestimated due to the presence of high optical depth HI which is not fully measured
in emission line observations.
The high optical depth HI can be measured from self-absorption features,
caused by the background Galactic HI emission being absorbed by the cooler
foreground HI
\citep{Knapp74, Goodman94, Li03}.
Many narrow self-absorption features have been considered as kinematically associated 
with CO and have inferred temperature of less than 40 K and 
the atomic hydrogen column density fraction
of only 0.0016 relative to H$_2$. If
HI is a dissociation product of H$_2$, these measurements suggest a cloud age of 3--30 Myrs 
\cite{Goldsmith05}.
While self-absorption can provide spatial information about the cold HI, 
e.g. \cite{GibsonS00}, it always requires complicated line modeling and is limited
by the ability to clearly distinguish self-absorption features from temperature fluctuations
and/or multiple individual line of sight components.

The main aim of this study is to investigate the effect of high optical depth on the HI surface
density saturation observed in Lee at al. (2012).
We use the most direct way to estimate the ``true'' HI column density by measuring
HI absorption against radio continuum sources located behind Perseus.
We use these observations to investigate properties of the cold and warm HI around Perseus
(Paper I), as well as to derive the
ratio of the true HI column density to the HI column density derived under the optically thin assumption (Paper II).

The structure of this study is organized in the following way.
In this paper (Paper I) we focus on the properties of cold gas around Perseus. 
Our observing and data processing strategies are explained in Section~\ref{s:obs}, and
in Section~\ref{s:analysis} we summarize the methodology used to estimate spin temperature and
column density of the cold neutral medium (CNM) and the warm neutral medium (WNM).
In Section~\ref{s:properties} we investigate the basic physical properties of atomic gas
in the Perseus HI envelope, and in Section~\ref{s:co_HI_abs} we
compare HI absorption and carbon monoxide (CO) emission spectra. 
We summarize our results in Section 6.  
In Paper II we estimate the correction for high optical depth
using our HI absorption measurements, apply this correction 
and re-visit the question of HI saturation in Perseus.

\section{Observations and Data Reduction}
\label{s:obs}

\begin{table*}
\begin{center}
{\bf TABLE 1} \\
\textsc{Source list} \\
\vskip 0.2cm
\begin{tabular}{ccccc} \hline \hline
Source & RA (J2000) & Dec (J2000)             & $T_{src}$ & $T_{sky}$  \\
        &    (h m s) & ($^{\circ}$ $'$ $''$) & (Jy)     &  (K) \\
\hline
NV0157+28 & 01:57:12.85 & 28:51:38.49 & 1.4 & 2.782 \\
4C+29.05 & 02:01:35.91 & 29:33:44.18 & 1.2 & 2.785 \\
4C+27.07 & 02:17:01.89 & 28:04:59.12 & 1.0 & 2.785 \\
5C06.237 & 02:20:48.06 & 32:41:06.64 & 0.9 & 2.787 \\
B20218+35 & 02:21:05.48 & 35:56:13.91 & 1.7 & 2.790 \\
3C067 & 02:24:12.31 & 27:50:11.69 & 3.0 & 2.786 \\
4C+34.07 & 02:26:10.34 & 34:21:30.45 & 2.9 & 2.791 \\
NV0232+34 & 02:32:28.72 & 34:24:06.08 & 2.6 & 2.791 \\
3C068.2 & 02:34:23.87 & 31:34:17.62 & 1.0 & 2.787 \\
4C+28.06 & 02:35:35.41 & 29:08:57.73 & 1.3 & 2.788 \\
4C+28.07 & 02:37:52.42 & 28:48:09.16 & 2.2 & 2.790 \\
4C+34.09 & 03:01:42.38 & 35:12:20.84 & 1.9 & 2.794 \\
4C+30.04 & 03:11:35.19 & 30:43:20.62 & 1.0 & 2.792 \\
B20326+27 & 03:29:57.69 & 27:56:15.64 & 1.3 & 2.787 \\
4C+32.14 & 03:36:30.12 & 32:18:29.47 & 2.7 & 2.793 \\
3C092 & 03:40:08.55 & 32:09:02.32 & 1.6 & 2.791 \\
3C093.1 & 03:48:46.93 & 33:53:15.41 & 2.4 & 2.795 \\
4C+26.12 & 03:52:04.36 & 26:24:18.11 & 1.4 & 2.783 \\
B20400+25 & 04:03:05.61 & 26:00:01.61 & 0.9 & 2.785 \\
3C108 & 04:12:43.69 & 23:05:05.53 & 1.5 & 2.788 \\
B20411+34 & 04:14:37.28 & 34:18:51.31 & 1.9 & 2.793 \\
4C+25.14 & 04:20:49.30 & 25:26:27.63 & 1.0 & 2.785 \\
4C+33.10 & 04:47:08.90 & 33:27:46.85 & 1.2 & 2.799 \\
3C131 & 04:53:23.35 & 31:29:25.36 & 2.9 & 2.801 \\
3C132 & 04:56:43.08 & 22:49:22.27 & 3.4 & 2.795 \\
4C+27.14 & 04:59:56.10 & 27:06:02.19 & 0.9 & 2.796 \\
3C133 & 05:02:58.51 & 25:16:25.16 & 5.8 & 2.796 \\
\hline
\end{tabular}
\end{center}
\end{table*}

\subsection{HI absorption observations}

\begin{figure*}
\centering
\includegraphics[scale=0.98,angle=0]{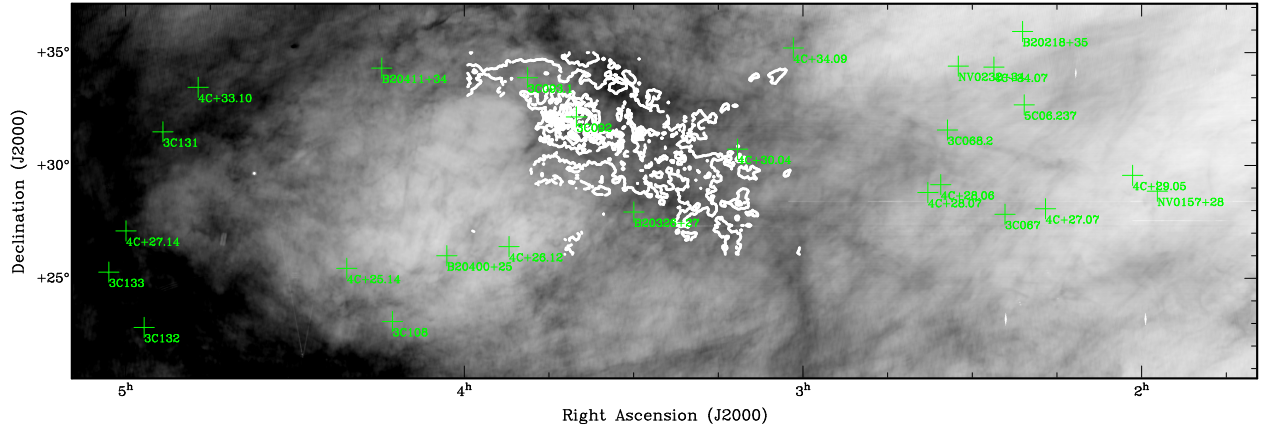}
\vspace{-2.5cm}
\caption{\label{f:srcs} Positions of background radio continuum sources overlaid on the
HI column density produced using GALFA-HI data at angular resolution of $\sim4'$. The intensity scale ranges
from $2.2 \times 10^{19}$ cm$^{-2}$ to $1.5 \times 10^{21}$ cm$^{-2}$.
White contours show the H$_2$ surface density
distribution of Perseus from Lee et al. (2012). Contour levels range from 5 to 90\% of the peak value ($4.6\times 10^{21}$
cm$^{-2}$), with a step of 10\%.}
\end{figure*}

We selected 27 radio continuum sources from the NVSS survey \citep{Condon98},
located over an area of roughly 500 square degrees centered
on Perseus with flux densities at 1.4 GHz greater than 0.8 Jy.
Figure~\ref{f:srcs} shows the source positions overlaid on 
the H$_2$ surface density image of Perseus from Lee et al. (2012).
Source information (RA, Dec, flux density at 21 cm, and the diffuse
background radio continuum emission) is given in Table 1.

The observations were conducted with  the Arecibo
telescope\footnote{
The Arecibo Observatory is operated by SRI International under a
cooperative agreement with the National Science Foundation (AST-1100968),
and in alliance with Ana G.
M\'{e}ndez-Universidad Metropolitana, and the Universities Space Research Association.}.
Using the L-wide receiver,
we simultaneously recorded spectra centered at 1420 MHz and
the two OH main lines (1665 and 1667 MHz), achieving a velocity resolution of 0.16 \kms.
We sampled simultaneously two linearly polarized channels performing both
auto and cross-correlations with the Arecibo's three-level ``interim" digital correlator.
The Arecibo telescope has an angular resolution of 3.5$'$ at these frequencies.
As shown by Heiles \& Troland (2003a) in their Millennium HI survey, 
Arecibo can accurately measure HI absorption lines for strong sources (flux density larger than
$\sim1$ Jy). 

The observing procedure used was the same as in \cite{Heiles03a} and
\cite{Stanimirovic05}.
This technique employs a 17-point observing pattern including 16 off-source
measurements and one on-source measurement.
The pattern was designed to measure the first and second derivatives of the 21-cm
intensity fluctuations on the sky, and also to fine-tune for the instrumental effects
involving the system gain.
The auto-correlation data were used to derive the ``expected" HI emission profile ($T_{exp}$),
which is the profile that would be observed at the source position if
the continuum sources were absent, the optical depth profile ($\tau$), and their
uncertainties. 
With 17 measurements, the off-source spectra are expressed
in a Taylor series expansion of the expected profile and a small contribution
from the source intensity attenuated by the optical depth.
A least-squares fitting technique
is then used to estimate the optical depth profile, 
the expected profile and its spatial derivatives, and
the off-source gain simultaneously \citep{Heiles03a}. 
However, our updated data reduction software takes a slightly
simpler approach by not including the fine-tuning of
gain variations under the assumption that the on-axis 
telescope gain and the beam properties vary spatially and a detailed
knowledge of these variations is required to estimate properly off-axis gains. Therefore,
we just derive the optical depth profile, the expected profile and its 
spatial derivatives for each of 16 off positions. 
These are used to derive the uncertainty spectra for both
the expected emission and optical depth spectra.

We have experimented with using the first order Taylor expansion instead of
the second order. For all sources we find that the difference between optical depth
profiles derived used the two expansions is within 1-$\sigma$ uncertainty.
While the second order expansion is clearly more accurate (has smaller systematic
errors), the derived $T_{exp}(v)$ and $\tau(v)$ are noiser than when 
using the first-order expansion. The increased noise comes from
fitting a larger number of unknown parameters, and also from a large covariance
between the second derivatives of the expected profile, $T_{exp}(v)$, and $\tau(v)$.
We tolerate the slightly higher noise for better accuracy of derived profiles
and therefore use the second-order Taylor expansion for all sources.

Following the data reduction, for all
sources  we obtained an HI absorption spectrum ($e^{-\tau(v)}$), an HI
(expected) emission spectrum ($T_{\rm exp}(v)$), and their uncertainty profiles.
A main beam efficiency of $\eta=0.85$ (based on calibration measurements
at Arecibo, Perillat et al.) was used to convert $T_{\rm exp}(v)$
from the antenna temperature units to the brightness temperature scale.
With an integration time on average of about 1 hour,
we achieved an rms noise level in optical depth of 
$\sim 1 \times 10^{-3}$ per 1 \kms velocity channel.

Inspection of derived profiles revealed that several sources have  
small positive spectral features in their optical depth profiles at a level slightly 
higher than the 1-$\sigma$ uncertainty and highly localized in velocity.  
This effect is a result of high spatial derivatives of the HI emission (due to the 
presence of significant small-scale structure) and suggests that even the second-order 
Taylor expansion is not a good representation of the measured off positions in 
several cases. These sources are: 4C+27.14 and 4C+33.10. 
In addition, 4C+33.10 has very broad both absorption and emission profiles with 
many velocity components and its component fitting is more difficult and ambiguous
than for other sources.
However, in order to use as many sources as possible and considering that
small artifacts are very localized in velocity,
we include these three sources in our analysis (but make sure that artifacts are
not fitted as real features). 
One source that we exclude from analysis is 4C+32.14 which has a 
highly saturated absorption profile 
and therefore all fitted parameters are highly uncertain for this source.

\subsubsection{Comparison with HT03}
Several of our sources were observed previously by \cite{Heiles03b} (from now
on HT03): 3C+93.1,
3C131, 3C132, and 3C133.
In terms of optical depth spectra, our results for 3C+93.1, 3C132, and 3C133
agree extremely well with HT03, within 3\%. In the case of 3C131 we find a slightly
larger difference, but this is still within the 3-$\sigma$ uncertainty.
In case of expected profiles expressed in terms
of antenna temperature, for all sources we find excellent agreement with HT03.
We do correct our expected profiles for the beam efficiency and work with
brightness temperature profiles in this paper.

\subsection{HI emission data from the GALFA-HI survey}

To investigate different methods for the derivation of the correction for high optical depth (focus of Paper II),
as well as to estimate the importance of stray radiation, 
we also use the H\textsc{i} emission data from the Galactic Arecibo L-band Feed
Array Survey in H\textsc{i} (GALFA--H\textsc{i}).
GALFA--H\textsc{i} uses ALFA, a seven-beam array of receivers
mounted at the focal plane of the 305-m Arecibo telescope, to map H\textsc{i} emission in the Galaxy.
Each of seven dual polarization beams has an effective beamsize of 3.9$'$ $\times$ 4.1$'$
and a gain of 8.5--11 Jy K$^{-1}$ \citep{Peek11}.
The GALFA--H\textsc{i} spectrometer, GALSPECT, has a velocity resolution of 0.184 km s$^{-1}$ (872 Hz)
and covers $-700$ km s$^{-1}$  $< v <$ $+700$ km s$^{-1}$ (7 MHz) 
in the Local Standard of Rest (LSR)
frame\footnote{All velocities quoted in this paper are in the kinematic or standard LSR frame, defined 
based on the average velocity of stars in the Solar neighborhood as:
20.0 \kms toward RA=18.0 hr, Dec=30.0 degrees in the 1900 epoch.}.

In Lee et al. (2012) we
combined scans from several GALFA-HI projects and produced an H\textsc{i} cube of Perseus centered
at (RA,Dec) = (03$^{\rm h}$29$^{\rm m}$52$^{\rm s}$,$+$30$^{\rm \circ}$34$'$1$''$)
in J2000\footnote{All quoted coordinates
in this paper are in J2000.} with a size of 14.8$^{\rm \circ}$ $\times$ 9.0$^{\rm \circ}$.
We use the same data here, but extend the data cube beyond Perseus to
include locations of all radio continuum sources.
This data cube has a size close to 60$^{\rm \circ}$ $\times$ 18$^{\rm \circ}$,
with a pixel size of 1$'$.
After smoothing the cube to 36$'$ and comparing the average HI spectrum with the
corresponding spectrum from
the Leiden/Argentine/Bonn (LAB) survey \citep{Kalberla05}, 
we derived the correction factor of 1.1
that needed to be applied on the pixel-by-pixel basis to fine-tune GALFA-HI's calibration (we note
that our data came from an early data reduction scheme, before the public GALFA-HI data cubes were 
finalized and released). 

Lee et al. (2012) also used the GALFA-HI data to investigate the HI saturation in Perseus. 
To estimate the HI column density, the  H\textsc{i} emission was integrated 
from $v_{\rm LSR} = - 5$ to 15 km s$^{-1}$. This range was selected as resulting 
in the maximum correlation between $N$(H\textsc{i}) and 
the A$_V$ image from 2MASS \citep{Ridge06}, exploring the idea that 
in mainly diffuse, low-A$_V$ regions of Perseus where molecular gas is not abundant
HI correlates well with A$_V$.

\subsection{Stray radiation consideration for HI emission}

Both our derived expected HI emission profiles and HI spectra
from the GALFA-HI survey may be affected by stray radiation.
Stray radiation is caused by radiation entering through higher order sidelobes
and can result in broad, weak emission features.
Correcting for stray radiation is a complex problem and requires a detailed
knowledge of the Arecibo telescope beam and how it varies with azimuth and elevation.
In this paper we provide only a rough check of our spectra relative
to the LAB survey, 
which has been meticulously corrected for stray radiation.
We take a twofold approach: (i) we compare our derived expected profiles
$T_{exp}$ with the HI spectra from the GALFA-HI survey and find good agreement
(within our estimated uncertainties);
(ii) we then smooth the GALFA-HI data cube to the same angular and velocity 
resolution of the LAB survey (36$'$),
extract spectra at the positions of our continuum sources and compare them to search
for broad wing-like features. We find that, in the majority of cases, the differences
lie below the 1-$\sigma$ uncertainty level for our derived expected profiles.
Therefore, we conclude that stray radiation is not a significant problem for this study.
Our future work will develop a methodology for a detailed stray radiation correction.

\begin{figure*}
\centering
\vspace{5pt}
\includegraphics[scale=0.8]{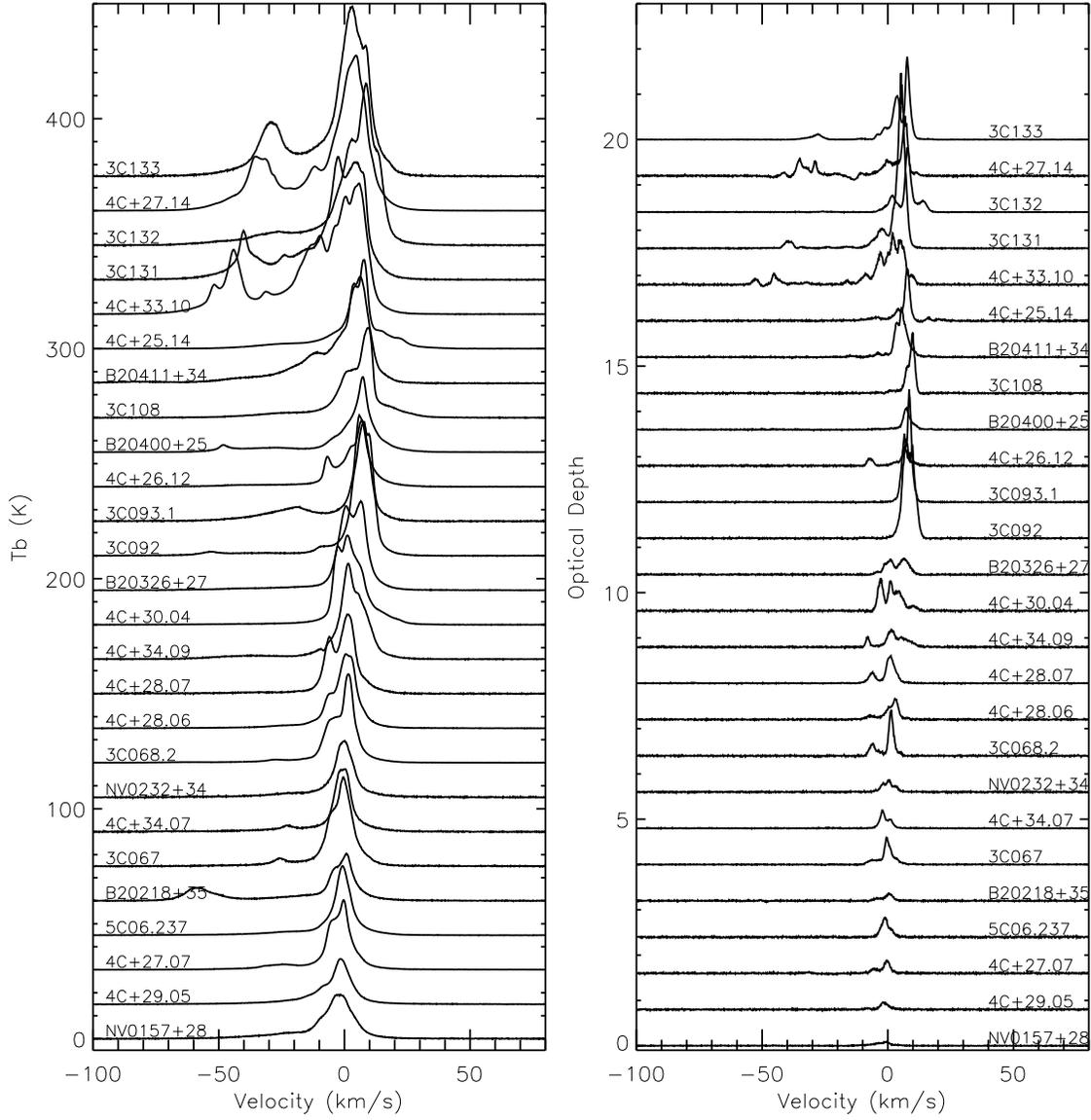}
\vspace{0pt}
\caption{(left) Brightness temperature of expected profiles of all sources offset in y-axis by 15 K for comparison.
(right) Optical depth profiles of all sources offset in y-axis by 0.8 for easy comparison.}
\label{f:all_src_emission}
\end{figure*}

\subsection{Additional data sets}

We use the CO (1-0) emission data from  \cite{Dame01} obtained with 
the 1.2 m telescope at the Harvard Smithsonian Center for Astrophysics (CfA) 
and at 8.4$'$ angular resolution.
We also use the integrated CO intensity ($W_{CO}$) and $E(B-V)$ images
from Planck \citep{Planck13} with angular resolution of 5$'$.
When using Planck data for comparison with \cite{Dame01} we first smooth the Planck 
images to angular resolution of 8$'$ and regrid to make sure pixels are independent.

\section{Analysis: Component fitting of HI absorption/emission pairs}
\label{s:analysis}

To analyze HI absorption spectra we performed a decomposition into
individual velocity components by
employing the technique of \cite{Heiles03a}. This allows us to estimate spin
temperature and the HI column density for individual CNM components. 
This technique assumes that the CNM
contributes to both HI absorption and emission spectra, while the warm
neutral medium (WNM) contributes only to the HI emission spectrum. The
technique is based on the Gaussian decomposition of both absorption and
emission spectra, and it takes into account the fact that a certain fraction
of the WNM gas may be located in front of the CNM clouds, resulting in a
portion of the WNM being absorbed by the CNM.
All possible permutations of the CNM components along the line-of-sight have
been taken into account when searching for the best fit. 
Pros and cons regarding the use of Gaussian functions to represent the CNM absorption profiles
have been discussed in \cite{Heiles03a}.

We first fit $\tau(v)$  with a set of $N$ Gaussian functions
using a least-squares technique:
\begin{equation}
\tau (v) = \sum_{0}^{N-1} \tau_{0,n} e^{-[ (v-v_{0,n})/\delta v_{n} ]^2}
\label{eqn:tau}
\end{equation}
where $\tau_{0,n}$ is the peak optical depth,
$v_{0,n}$ is the central velocity, and $\delta v_{n}$ is the 1/e width
of component $n$. $N$ is the minimum number of components necessary to make the residuals of the 
fit smaller or comparable to the estimated
noise level of  $\tau(v)$.

While the optical depth spectrum predominantly  reflects the CNM, both the cold {\it and} warm 
neutral media contribute to the expected HI emission spectrum:
\begin{equation}
T_{\rm exp}(v)= T_{B,CNM}(v) + T_{B,WNM}(v).
\label{eqn:Tb}
\end{equation}
The first term, $T_{B,CNM}(v)$, the HI emission originating from $N$ CNM components is:
\begin{equation}
T_{B,CNM}(v) = \sum_{0}^{N-1} T_{s,n}(1-e^{-\tau_{n}(v)})
e^{-\sum_{0}^{M-1} \tau_{m}(v)},
\label{eqn:Ts}
\end{equation}
where $T_{s,n}$ is the spin temperature of cloud $n$, and
the subscript $m$ represents each one of the $M$ CNM clouds that lie
in front of cloud $n$.

Next, $T_{B,WNM}(v)$, the HI emission originating from the
WNM, is represented with a set of $K$ Gaussian functions. The complicating factor here is that a certain
 fraction $F$ of the WNM is
located in front of the CNM, while a fraction $(1-F)$
of the WNM is beyond the CNM with its emissions being absorbed by CNM clouds: 
\begin{equation}
T_{B,WNM}(v)= \sum_{0}^{K-1} [F_{k} + (1-F_{k})e^{-\tau(v)}] \times
T_{0,k}e^{-[(v-v_{0,k})/\delta v_{k}]^2 },
\end{equation}
where the subscript $k$ corresponds to each of the WNM components and a fraction
$F_{k}$ of the WNM cloud $k$ lies in front of all CNM components, while a
fraction $1-F_{k}$ is being absorbed by the CNM clouds.
To fit the corresponding emission spectra, we assume that the center and width of the 
absorption-selected CNM components are fixed and include a minimum number of 
additional WNM components to reduce the fit residuals to within the neighborhood of the 1-$\sigma$ uncertainties.
We use a certain number of WNM components and fit the $T_{\rm exp}(v)$ profile 
simultaneously for the Gaussian
parameters of the WNM components and the spin temperature of individual
CNM clouds, while assuming a given order of CNM clouds along the 
line of sight and a given set of $F_{k}$ values.
We try to use the minimum number of WNM components such that the
residuals of this fitting process are reasonably close to the 1-$\sigma$
uncertainty for $T_{\rm exp}$.

Please note that the expected profile in the left-hand side of equation (2)
has been baseline corrected, which means that we measure
$T_{\rm exp}(v)-T_{sky}$, where $T_{sky}$ contains contributions from the Cosmic Microwave Background (CMB) 
and the Galactic synchrotron emission. Before doing the radiative transfer calculations 
we estimate $T_{sky}$ and add it back
to the left-hand side of equation (2) by assuming 2.725 K for the CMB.
To estimate the contribution from the Galactic synchrotron emission we use the \cite{Haslam82} 408 MHz
survey of the Galaxy. The brightness temperature at 408 MHz is converted to 1.4 GHz using the spectral
index of $-2.7$. As the Galactic latitude of observed sources in the vicinity of Perseus is generally $>10$ degrees,
the synchrotron contribution is small and $T_{sky}$ ranges 
from 2.78 to 2.80 K in our case (Table 1).

For each source, we vary the order of Gaussian functions along the
line-of-sight (for $N$ CNM components there are $N!$ possible orderings)
and perform the $T_{\rm exp}(v)$ fit. We then choose the ordering of CNM components that gives the smallest
residuals in the least-squares fit.
Unfortunately, the
difference in the fit residuals is often not sufficiently statistically significant to
distinguish between different values of $F_{k}$. However, $F_{k}$ has
a large effect on the derived spin temperatures. Hence we follow the
\cite{Heiles03a} suggestion and estimate the final spin temperatures by
assigning characteristic values of 0, 0.5, or 1  to each $F_{k}$ (among the extreme possible 
values of 0 and 1), and repeating this for all possible
combinations of WNM clouds. The final spin temperatures are then
derived as a weighted average over all trials.

Out of 26 sources, 23 have well-constrained fits. 
Three sources, 3C133, 3C131 and 4C+25.14, have more than 6 individual CNM components 
in their absorption spectra. The corresponding fit for the spin temperature of these components 
in the presence of WNM features in emission is therefore more complicated, 
and the fitting process does not converge. Furthermore, for 6 sources (3C068.2, 
3C133, 4C+25.14, 4C+28.07, 4C+30.04, and B20411+34) the fitted height of 
one absorption component is too small to be reliably recovered in the 
corresponding emission spectrum. Thus, the spin temperatures for these 6 components are calculated to 
be less than 1 K. 
Increasing the spin temperature by hand does not significantly degrade the quality of 
the fit. Therefore, for these uncertain components, we set the spin temperature 
equal to the uncertainty in $T_s$ derived from the iterations over CNM component 
orders along the line of sight and fraction of WNM absorbed. 
The error on this value is set to the median $T_s$ error for components along all 
26 lines of sight, or 6.75 K.

\begin{figure*}[b!]
\centering
\includegraphics[width=0.47\textwidth]{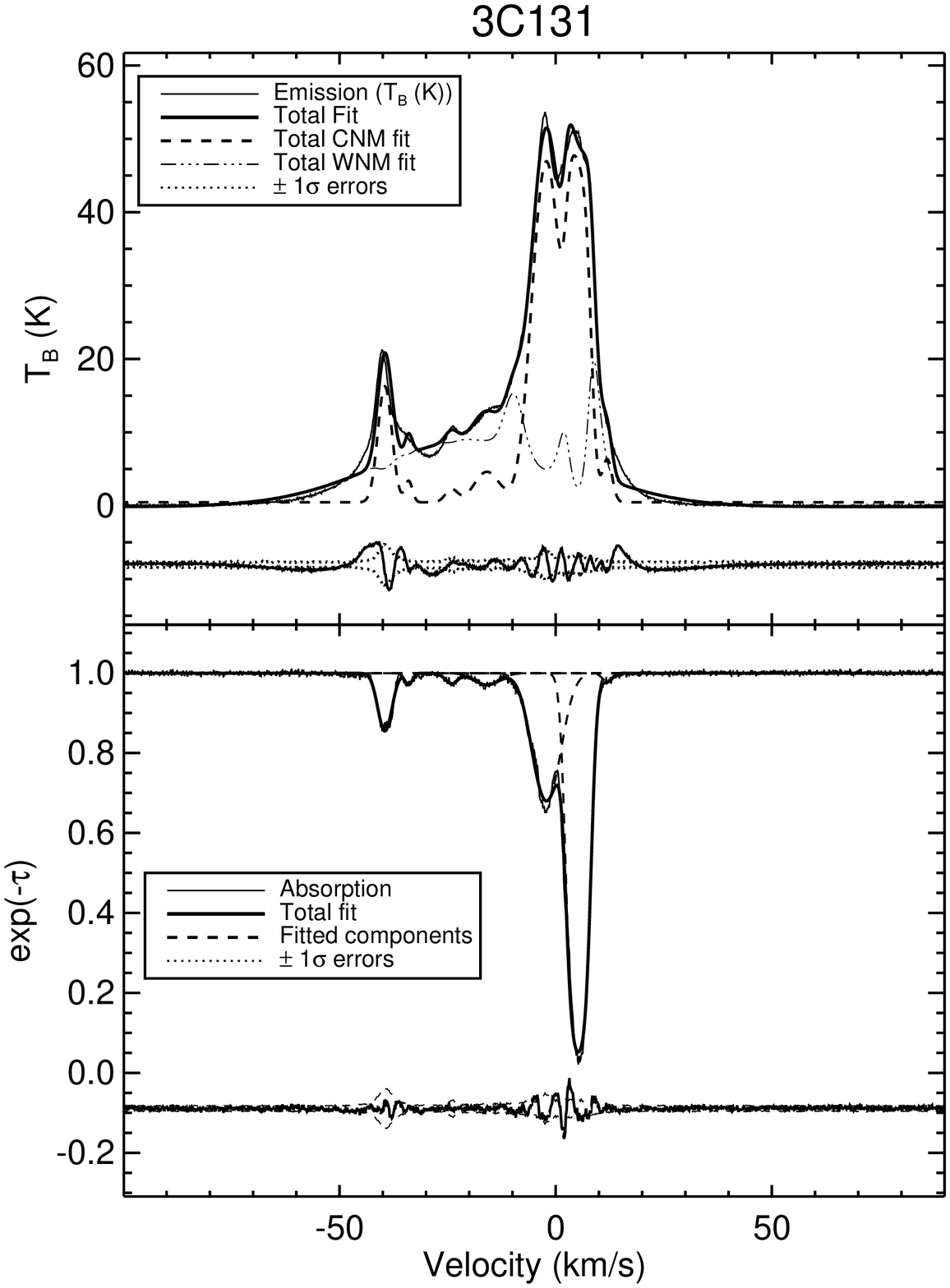}
\includegraphics[width=0.47\textwidth]{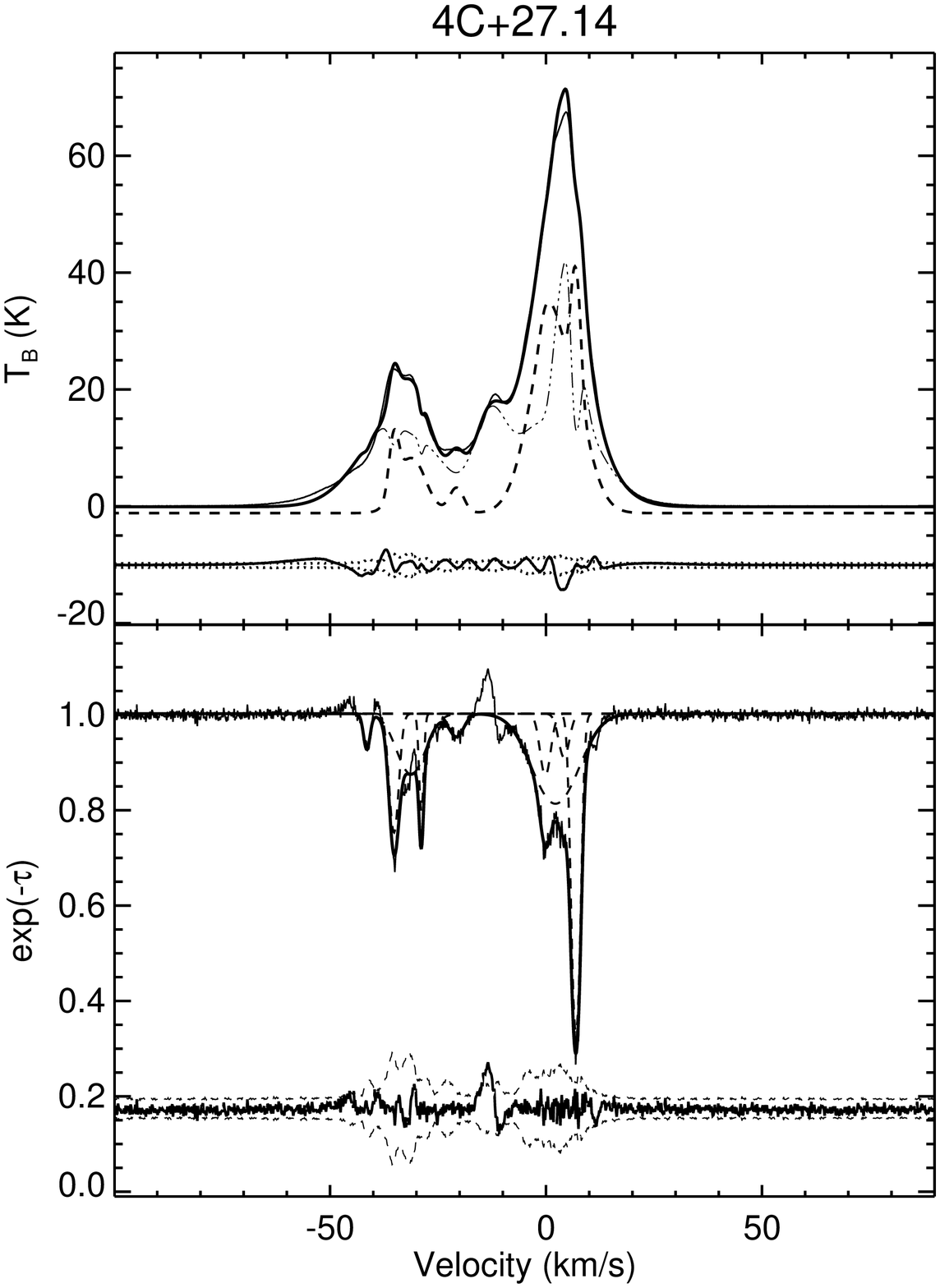}
\vspace{0pt}
\caption{Example Gaussian fits to emission and absorption spectra. (Left) 3C131,
(Right) 4C+27.14. In the top panels, the thin solid line is the expected profile, $T_{\mathrm{exp}}$ 
(see Section 2.1 for derivation). The thin dot-dashed lines display the sum  of WNM Gaussian components 
and the thick dashed lines display the total contribution to the $T_{\mathrm{exp}}$ profile by the 
CNM from the absorption profile. The thick solid line is the total WNM and CNM fit. 
The residuals from the fit are plotted below zero, with +/- $\Delta T_{\mathrm{exp}}$ overplotted. 
In the bottom panel, the thin solid line shows the optical depth 
profile ($e^{-\tau}$), with CNM 
components displayed in the thin dotted lines and the thick 
solid line representing the fit to
the optical depth profile. The residuals from the fit 
are plotted at the bottom of the figure, 
with +/- $\Delta e^{-\tau}$ overplotted. }
\label{highfraction}
\end{figure*}

\begin{figure*}[b!]
\centering
\includegraphics[width=0.47\textwidth]{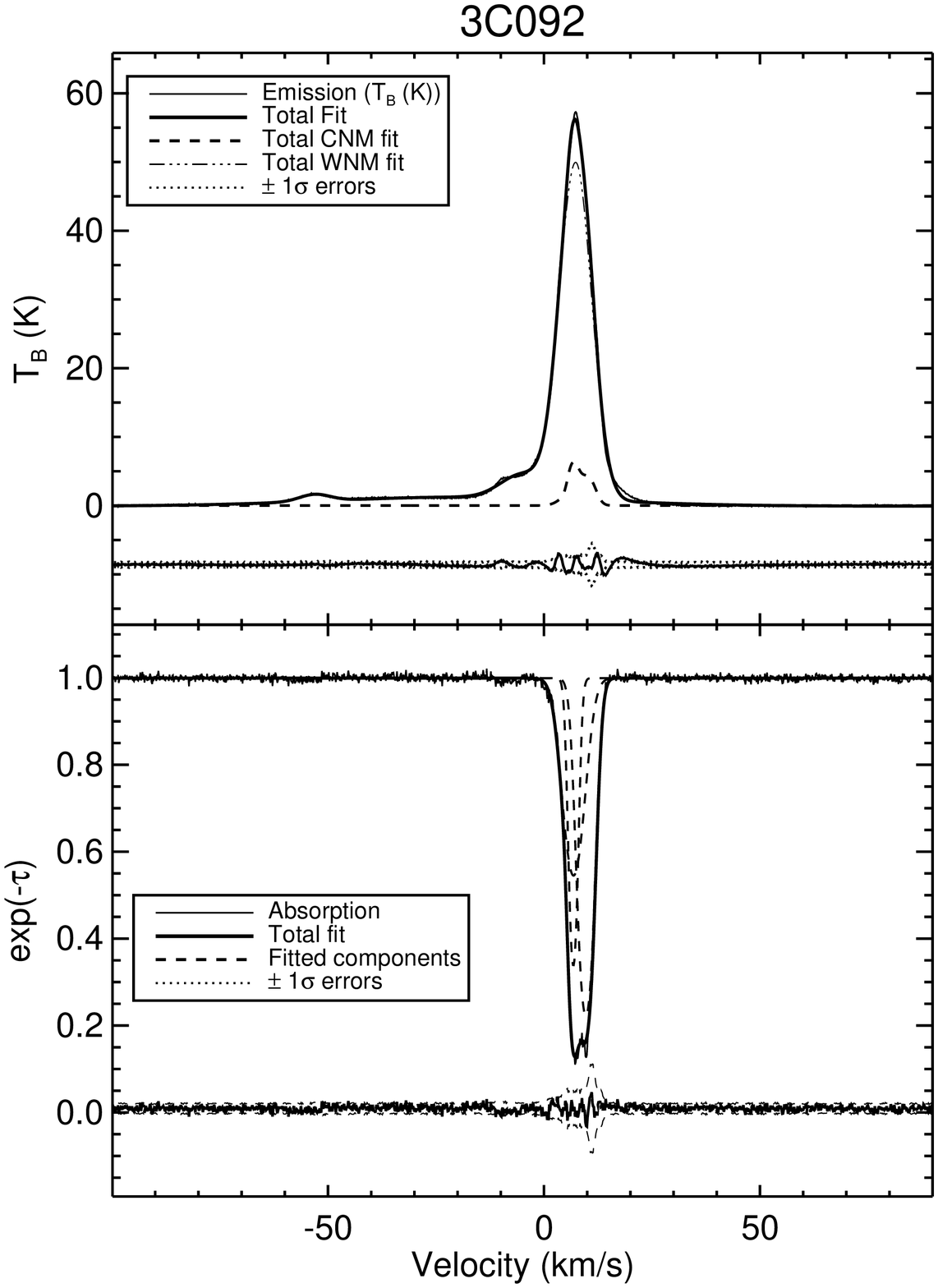}
\includegraphics[width=0.47\textwidth]{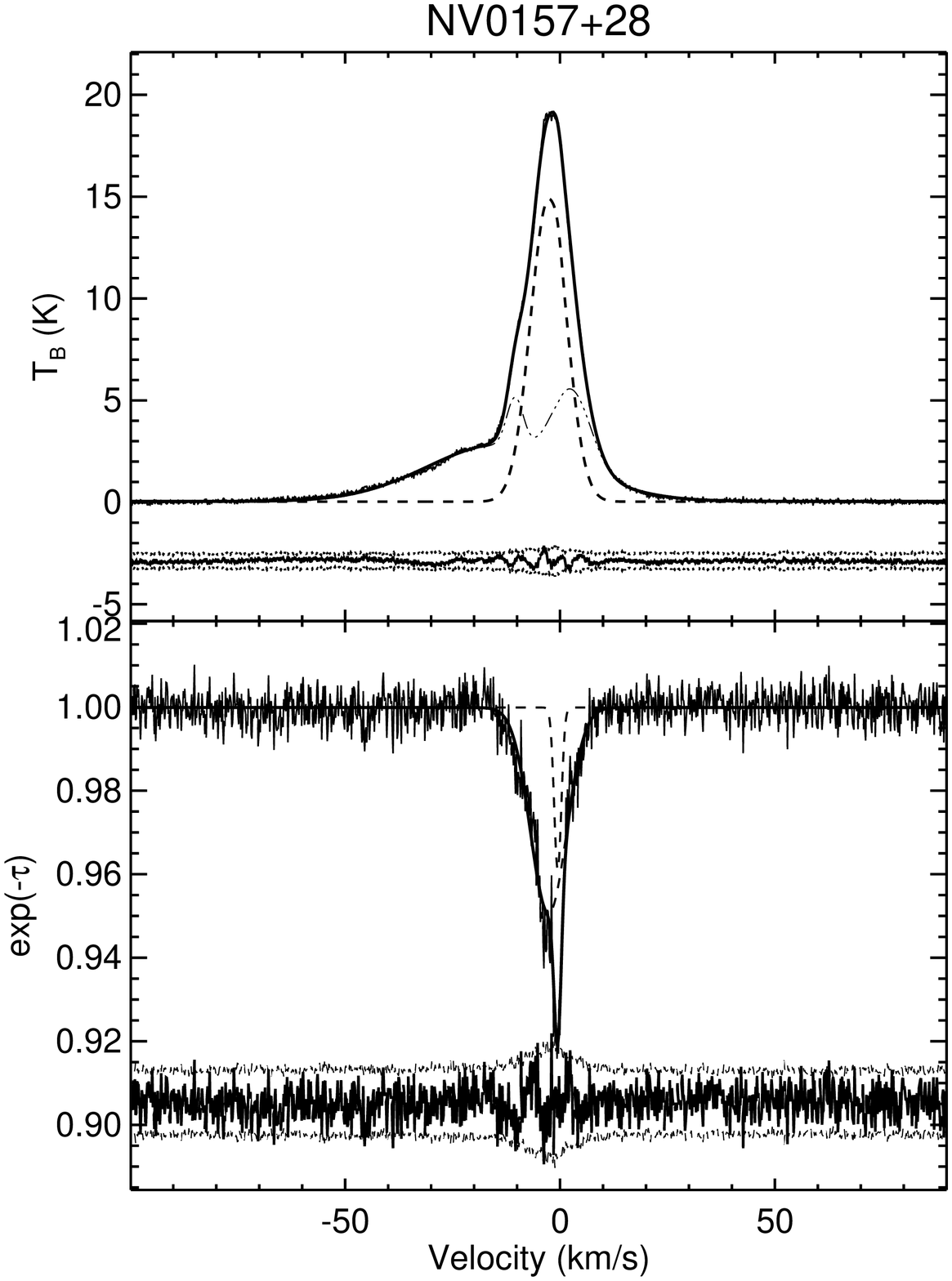}
\vspace{0pt}
\caption{Example Gaussian fits to emission and absorption spectra. 
(Left) 3C092 which is located behind the main body of Perseus,
(Right) NV0157+28. See Figure~\ref{highfraction} for a detailed description of the panels.}
\label{lowfraction}
\end{figure*}

\section{Properties of cold and warm gas around Perseus}
\label{s:properties}

Figure~\ref{f:all_src_emission} shows emission and absorption 
spectra for all sources except
4C+32.14 which has a saturated optical depth profile.
Strong absorption lines were detected in the direction of all sources.
In all cases, the strongest emission and absorption is found at $\sim0$ \kms,
and is generally well confined within the range of $-20$ to 20 \kms (see also Figure 6).
However, in the case of four sources (4C+33.10, 4C+27.14, 3C133, and 3C131)
there are strong emission and absorption features around $-40$ \kms.
Visual inspection of velocity components close to Perseus 
using the GALFA-HI data cubes
suggests that this
secondary region is likely not associated with Perseus.

We show results of our Gaussian component fitting for 4 example sources in 
 Figures \ref{highfraction} and \ref{lowfraction}. 
In each panel of both figures we plot the derived expected emission and optical depth 
profiles for an individual source. For the optical depth spectra, 
we overplot the individual CNM components (dotted lines), as well as 
the residuals for the fit (offset to the bottom of the panel) 
with the derived uncertainties for the spectrum  
for comparison. For the expected emission profiles, we overplot the sum of all 
WNM components (dot-dashed line), the total $T_s$-corrected contribution of the CNM 
(thick dashed line), and the fit residuals (shown below zero in 
the panel) with the uncertainties in the profile for comparison. 
The two sources in Figure \ref{highfraction}, 3C131 and 4C+27.14, have 
 broad HI profiles as likely include emission/absorption beyond Perseus,
3C092 in Figure \ref{lowfraction} is located behind the main body of Perseus, 
and NVO0157+26 in the same figure is an example of a low optical depth profile.

In Table 2, we list the Gaussian parameters associated with all CNM and WNM 
components for each source. In column 1 we list the peak brightness temperature 
for each component. For the WNM components, this is equal to the unabsorbed Gaussian 
height and estimated error in the fit. For the CNM components, this is equal to 
the calculated spin temperature multiplied by (1-$e^{-\tau}$), as in Equation~\ref{eqn:Ts}, 
and is quoted without uncertainty. In columns 2 and 3 we list the centers and FWHM of 
CNM and WNM components with estimated fit uncertainties. In column 4, we list the 
peak optical depth of each component. For the CNM components, this is equal to the height 
of each component (in $\tau$), with associated uncertainty. For the WNM components, 
this is equal to the maximum contribution of each WNM component detected in emission 
to the absorption profile, and is found by measuring the height of the absorption fit 
residuals at the central velocity of each WNM component. In column 5 we list 
the spin temperatures, which for the CNM components is equal to the calculated values 
from the fit   with fit uncertainties. For the WNM components, this is equal to a 
lower limit imposed by the upper limit on optical depth in column 4, and these values are 
also quoted without error because the errors are extremely large due to the nature of the 
estimation process. In column 6, we list the maximum kinetic temperature of each component 
based on the line widths. In column 7, we list the HI column density of each individual 
component, and these values are quoted in units of 10$^{20}$ cm$^{-2}$. Finally, in column 8 
we list the fraction of each WNM component lying in front of all CNM components ($F$, either 
0.0, 0.5 or 1.0, see Section 2.2) or the order of each CNM component along the line of sight ($O$, integer values).  

\subsection{Optical Depth}

\begin{figure*}
\centering
\vspace{5pt}
\includegraphics[width=\textwidth]{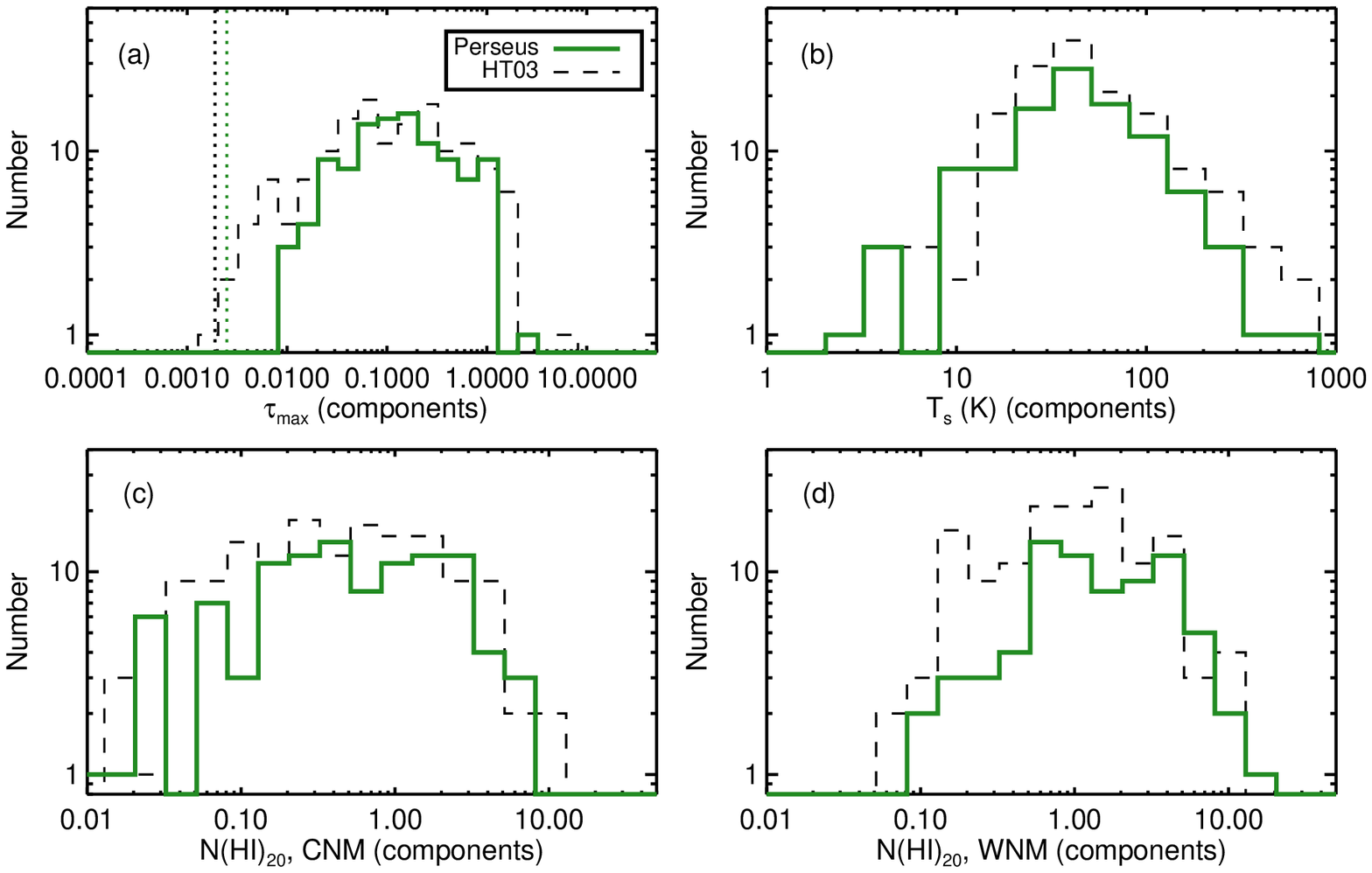}
\vspace{-275pt}
\caption{Histograms of Gaussian fit parameters for individual Gaussian 
components: (a) peak optical depth $\tau_{\rm max}$,
(b) spin temperature, (c) the CNM column density 
(in units of $10^{20}$ cm$^{-2}$), and (d) the WNM column density 
(in units of $10^{20}$ cm$^{-2}$).
Gaussian components from the HT03 survey at $|b|>10$ degrees are shown with a dashed  black line for comparison.
Dotted lines show the median sensitivity in optical depth for two studies. 
We assume here as the CNM all HI detected in absorption.}
\label{f:tau-Ts}
\end{figure*}

A summary of the fitting results is presented in Figures~\ref{f:tau-Ts} to \ref{f:dist_fract}. 
As shown in Figure~\ref{f:tau-Ts} (a),
the median peak optical depth $\tau_{max}$ for individual Gaussian components is 0.16, 
and only a handful of CNM components has $\tau_{max}>1$ (10/107).
Perseus is an intermediate-mass GMC located about 20 degrees below
the Galactic plane and may not sample the densest molecular gas. 
In addition, a tighter grid of background sources may be able to sample better 
denser gas. Only two of our sources are located right behind the main body of Perseus. 
Their peak optical depth is 1.5.

The same figure shows $\tau_{max}$ for the components from HT03, dotted lines show median
rms noise in optical depth for two studies.
The two studies agree very well and have relatively similar (median) sensitivity, but we
are missing the low-$\tau_{max}$ portion of the distribution.
This could be partially due to our small survey area relative to HT03 who 
had more sources at high Galactic latitudes.
We note that HT03's sensitivity varies across sources as their survey 
was searching for strong sources suitable for Zeeman measurements.

Very recently, \cite{Fukui14b} suggested a new approach to
estimate properties (optical depth and spin temperature) of
cold HI by utilizing dust emission. They noticed that 
the Planck dust optical depth $\tau_{353}$ at 353 $\mu$m correlates with $N(HI)$,
but the scatter in this relation is much smaller 
when different dust temperature regimes are considered separately.
By assuming that the highest dust temperature sub-sample is associated
with the optically-thin HI,
the saturation seen in the $\tau_{353}$-$N(HI)$ relation was 
attributed to the existence of the high optical depth HI solely.
By inverting the relation, they estimated a single value 
of $T_s$ and $\tau_{HI}$ per pixel from their all-sky $\tau_{353}$
images (after masking low-latitude regions with $|b|<15$ degrees and regions with
internal dust heating as traced by the H$\alpha$ emission).
They found that 85\% of data points have $\tau_{HI}>0.5$ and $T_s<40$ K.
Similar results were obtained for the high latitude clouds MBM 53-55 \citep{Fukui14a},
increasing the HI mass of MBM 53-55 clouds by a factor of two.
 \cite{Fukui14b}  suggested that the local
interstellar medium (ISM)
may be dominated by the high optical depth HI, and that this component 
may explain all of the CO-dark gas in the Milky Way.

Around Perseus we find $\tau_{max}>0.5$ only for 21 out of 107 (20\%) 
individual (Gaussian) components. This is clearly in stark contrast
with \cite{Fukui14b} who claimed that 85\% of lines of sight 
at essentially $|b|>15$ degrees have
$\tau>0.5$ based on their comparison of $\tau_{353}$ and $N(HI)$.

\subsection{Spin Temperature}

\begin{figure*}
\centering
\includegraphics[width=0.7\textwidth]{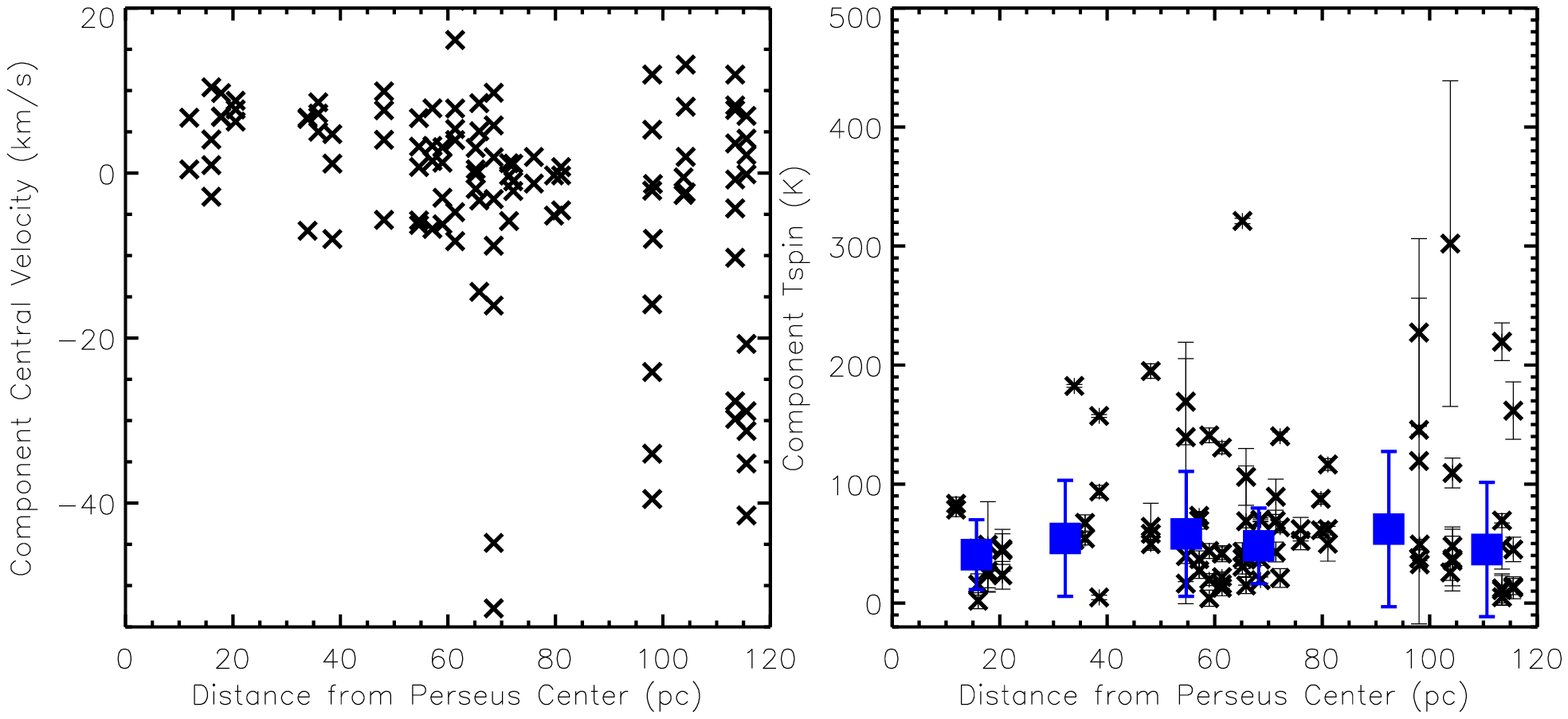}
\caption{(Left) The central velocity of all Gaussian components showing that
most components are found within $-20$ to 20  \kms~ velocity range.
(Right) The spin temperature of Gaussian components 
with velocity centroid within $-20$ and 20 \kms~ shown
as a function of distance from the Perseus center. Squares show
median values of $T_s$ calculated over 20 pc wide distance bins. 
The angular separation has been converted into
linear distance assuming a distance of 300 pc.}
\label{f:dist_Ts}
\end{figure*}

Figure~\ref{f:tau-Ts} (b) shows our estimated spin temperature which 
ranges from $\sim5$ to 725 K, with most
CNM components having $T_s=10$ to 200 K. 
The spin temperature distribution peaks at $\sim50$ K,
the median value is 49 K. This is in excellent agreement 
with HT03 results based on 66 random lines of sight at $|b|>10^{\circ}$,
as shown in the same figure.
While we have a slightly smaller number of components relative to the HT03 study,
the agreement between two studies is excellent over the full temperature range. 
In summary, the component spin temperature, for the predominantly CNM population
we are tracing in absorption, is similar between a large angular area and 
a more focused area around Perseus.

In addition to HT03,
one of our sources, 3C093.1, was observed by \cite{Andersson92} 
who found only one absorption component
and estimated its spin temperature of  41 K. The line of sight to 
this source pierces  through the main body of Perseus.
We have fitted the HI absorption spectrum with 
three components, and their spin temperature is
$45\pm13$, $44\pm17$ and $23\pm11$, respectively.
The range of spin temperature in the direction of additional seven sources observed by
Andersson, Roger, Wannier (1992) is $40-100$ K.
This all shows that our derived temperatures are in general agreement with previous studies.
Our mean $T_s$ is also in agreement with an estimate from Lee et al. (2012) of 60-75 K, where the equilibrium KMT09 model
for the H$_2$ fraction was fitted to observations, under the overarching assumption of the CNM and WNM co-existing 
in pressure equilibrium.

In stark contrast to \cite{Fukui14b}, we find 
the spin temperature distribution essentially identical to
an average CNM temperature distribution for the Milky Way, e.g. HT03 or 
\cite{Strasser07}.
Out of 107 absorption-detected Gaussian components, $\sim50$\% have $T_s<50$ K. 
There are three sources whose projected distance from the rough center of
Perseus is less than 20 pc, and their mean spin temperature is 45 K.
The low spin temperature (20-40 K for 85\% of data points) in Fukui et al. could be an artifact of
neglecting to account for the ``CO-dark'' H$_2$  gas in the $\tau_{353}$-N(HI) correlation,
and the use of line-of-sight averaged properties (single spin temperature and optical depth).

In Figure~\ref{f:dist_Ts} (left) we show the central velocity
of all Gaussian CNM components which shows that most components have a 
central velocity between $-20$ and 20 \kms. In Figure~\ref{f:dist_Ts} (right)
we plot $T_s$ for all components within this velocity range, excluding components
that are likely (based on their central velocity) not associated with Perseus.
Squares show median $T_s$ over 20 pc wide bins. We do not find obvious variations 
of $T_s$ with the distance from the center of Perseus.

Spatial changes of $T_s$ across interstellar clouds have been claimed in the literature.
\cite{Liljestrom88} mapped the HI distribution of a high latitude cloud
and interpreted the observed increase in the line width as being due to an increase
of $T_s$ by $\sim30$ K.
Andersson, Roger, Wannier (1992) performed radiative transfer modeling 
of HI observations of the B5 region in Perseus,
considering
internal stars and the effect of stellar winds on the spin temperature distribution.
Their model suggests spin temperature of 40-50 K within the first 2 pc from the central star cluster, 
and then an increase to 200-300 K out to 6-8 pc from the core center.
We do not find any evidence for a systematic change of $T_s$ radially from the Perseus center
as shown in Figure~\ref{f:dist_Ts}, however we have large gaps in the  background source coverage. 
A much tighter grid of HI absorption spectra within 50 pc from the center would be important for future studies.

\subsection{HI column density and the CNM fraction}

\begin{figure*}
\centering
\vspace{5pt}
\includegraphics[scale=0.9]{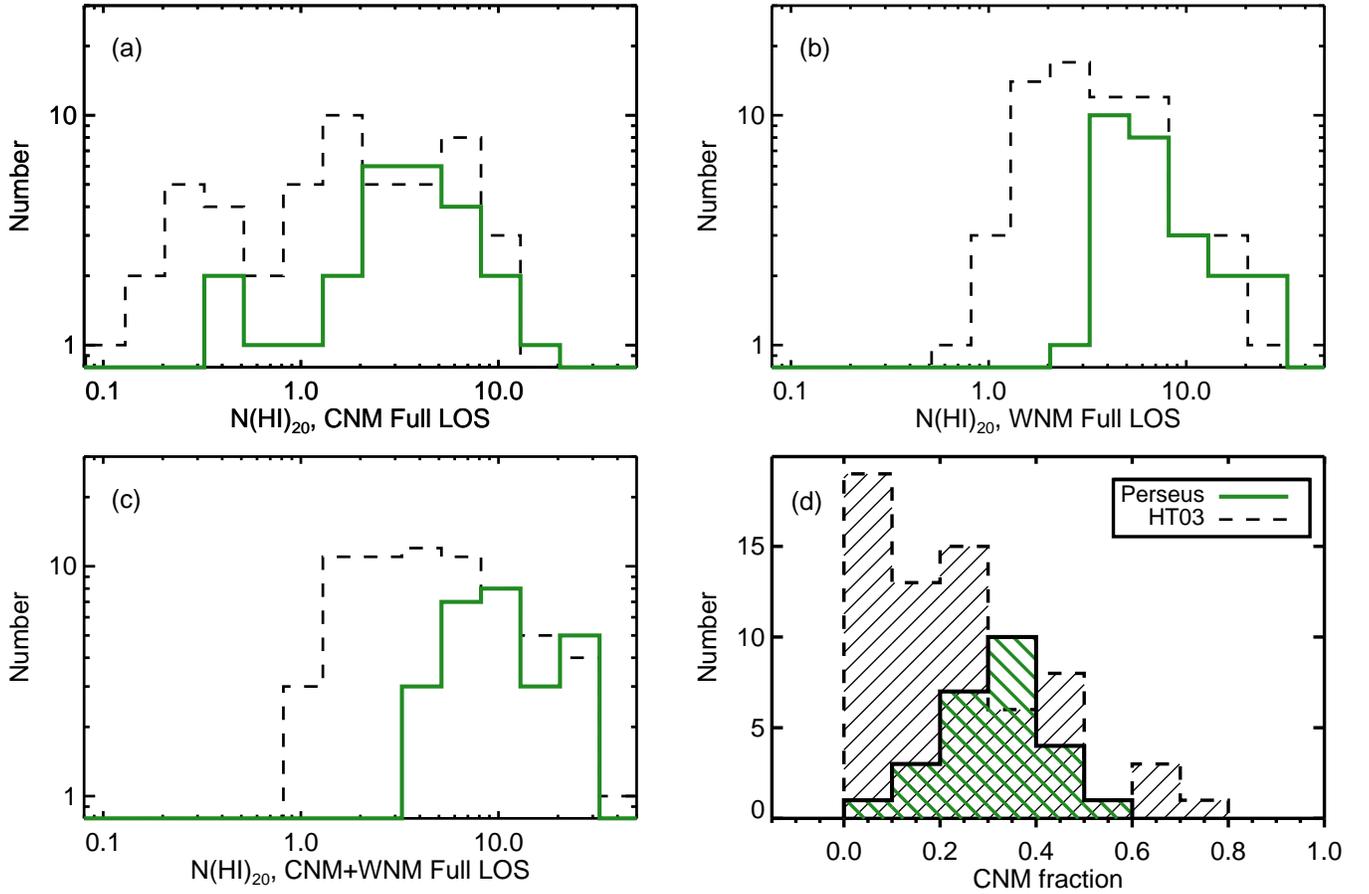}
\vspace{-300pt}
\caption{Histograms of basic properties calculated for each line of 
sight, from this study as solid green and from HT03 (for 
their $|b|>10^{\circ}$ sources)
as dashed black:
(from top left to bottom right) the CNM column density, 
the WNM column density, the CNM+WNM column density,
and the CNM fraction (CNM/(CNM+WNM) column density). 
A $T_s$ threshold of 200 K was applied when selecting CNM components.    }
\label{f:CNM_WNM_fullLOS}
\end{figure*}

Histograms of the CNM and WNM column densities derived for individual Gaussian 
components are shown in Figure~\ref{f:tau-Ts} (c) and (d) as solid 
green, while the results from the HT03 survey are
shown as dashed black. There is excellent agreement 
between two studies\footnote{To quantify this we have
calculated cumulative distribution functions for $\tau_{max}$, $T_s$, $N(HI)_{CNM}$ and $N(HI)_{WNM}$
to compare our results with HT03. The K-S test suggests that there is 83\% and 72\% 
probability that Perseus and HT03 $T_s$ and $N(HI)_{CNM}$ distributions were drawn from the same sample.}.
Our median CNM column density is $6.0 \times 10^{19}$ cm$^{-2}$, 
in comparison to $5.2 \times 10^{19}$ cm$^{-2}$ by HT03.
Our median WNM column density is $1.5 \times 10^{20}$ cm$^{-2}$, 
in good agreement with $1.3 \times 10^{20}$ cm$^{-2}$ estimated by HT03.
Please note that both studies treated as the CNM {\it all} HI detected in absorption and no
temperature selections were made to distinguish the CNM from the thermally unstable WNM.
It is interesting to note that Figure~\ref{f:tau-Ts} shows 
that the WNM has a more uniform column density, while the 
CNM column density varies more dramatically, from  $10^{18}$ to $10^{21}$ cm$^{-2}$.

In Figure~\ref{f:CNM_WNM_fullLOS} we show integrated CNM and WNM 
properties for different lines of sight probed by our target
background sources, as well as results from HT03 for their 66 random directions
at  $|b|>10^{\circ}$.
As several of our CNM components have higher temperature likely
more appropriate for the thermally unstable WNM (e.g. Kim et al. 2014),
we have applied a temperature threshold of $T_s<200$ K 
when calculating the CNM column density and
the CNM fraction along the line of sight. The same threshold was applied for the HT03 data.

The main conclusion from this figure is that the line of sight properties 
in our study trace the upper range of the HT03 histograms. 
In terms of details, we find median CNM and WNM HI column density of 
$4.5 \times 10^{20}$ cm$^{-2}$ and $7.8 \times 10^{20}$ cm$^{-2}$, respectively.
Both values are more than five times higher than the corresponding median 
values in HT03. The same applies to the total HI column density.
To quantify the disagreement between our study and HT03 we have
calculated cumulative distribution functions for 
all distributions in Figure 7.
The K-S test suggests that there is $\la3$\% 
probability that Perseus and HT03 line-of-sight distributions were drawn from the same sample.

HT03 found a large number of sources with $\Sigma N(HI)_{CNM} =0$ as 
can be seen in Figure~\ref{f:CNM_WNM_fullLOS}(d) where $\sim15$
of HT03's sources did not have detectable CNM.
\cite{Stanimirovic05} and \cite{Stanimirovic07}
showed that with $>4$ times longer integrations
weak CNM features were detected in some of these directions.
For each of our 26 sources we detect significant HI absorption lines
with the CNM fraction being $>20$\% for 20 sources, 
the lowest CNM fraction we find is 1\% and 
there is only one source with such low fraction. 
As the sensitivity of two studies is on average similar, 
our higher fraction of absorbing HI 
likely stems from the intrinsic properties of the Perseus region.
Our median CNM fraction is 0.33, in comparison to
0.22 in HT03 (after the same 200 K cut-off was applied to both studies). 
The above results strongly suggest
that the Perseus region has a higher fraction of absorbing HI and a higher total HI 
column density relative to an average ISM field. 
The absorbing HI appears to contribute significantly to the 
total column density along almost every line of sight.

In summary, while properties of individual components are in 
excellent agreement with those of HT03,
it appears that the Perseus region has a larger number of 
absorbing HI components relative to an average, random 
ISM field. This could explain the enhanced total HI column density and 
the fraction of the absorbing HI.
Our results in particular for the CNM (and to a smaller degree for 
the WNM) and the total $N(HI)$ essentially trace the upper 
range of the corresponding distributions from HT03. 

The CNM fraction, and especially its variations with interstellar environments,
are poorly constrained observationally.
In a comprehensive study of 290 HI emission/absorption pairs,
\cite{Dickey09} showed that the radial dependence of the harmonic mean 
spin temperature, which is a product of the spin temperature and the CNM fraction, 
is flat across the Milky Way disk.
Considering that Strasser et al. (2007) showed that spin temperature of the CNM 
is similar between the inner and outer Galaxy,
this result implies a nearly constant CNM fraction out to 25 kpc. 
Our study of Perseus is the first hint that the CNM fraction 
in/around GMCs is likely higher than what is found in an average ISM field.

\subsection{What determines the CNM fraction?}

While the Perseus region has on average a higher CNM fraction relative to an
average ISM field, as shown in Figure~\ref{f:CNM_WNM_fullLOS},  
interestingly almost all directions (25/26)
in our study have the CNM fraction smaller than 50\%. 
We emphasize that this result is not an artifact of our applied $T_s<200$ K cutoff.
If we do not apply any temperature cutoff, the median CNM fraction is 35\%,
and 23/26 directions have a CNM fraction $<50$\%.

\begin{figure*}
\centering
\vspace{5pt}
\includegraphics[scale=0.8]{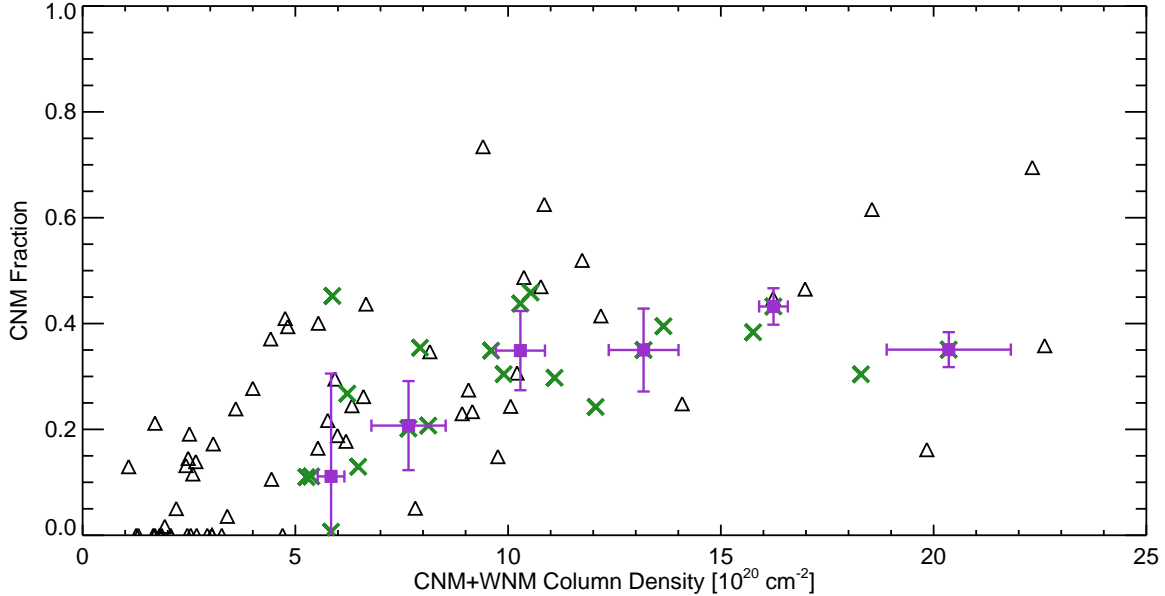}
\vspace{-350pt}
\caption{The CNM fraction as a function of the total HI column density,
green crosses show our data and black triangles are from HT03. Purple points show 
median values with 1-$\sigma$ scatter calculated for our 
observations only. To isolate CNM-only 
and exclude potentially thermally-unstable WNM we have applied
a cutoff $T_s<200$ K for both data sets.}
\label{f:dist_fract}
\end{figure*}

In Figure~\ref{f:dist_fract} we show the CNM fraction as a function
of the total HI column density for our sources as well as HT03 data.
It is obvious that the CNM fraction 
never gets higher than $\sim80$\% (for both studies). 
This shows that there are no lines of sight without the WNM,
even in the directions where the CNM hugely dominates
the WNM fraction is at least $\sim20$\%.
Although the scatter in this figure is large, the CNM fraction appears to
increase from 0 to $\sim40$\% at $N(HI) \sim 10^{21}$ cm$^{-2}$, and then 
levels off (purple points show median values for Perseus observations).
As pointed out by \cite{Heiles03b}, this transition 
occurs right around the column density
required for shielding H$_2$, suggesting that
the CNM transitions into H$_2$ as soon as the adequate shielding is achieved.
This column density also agrees with Lee et al. (2012) who showed that
the HI-to-H$_2$ transition (defined as having a 
H$_2$ fraction of 0.25) occurs in Perseus at
$N(HI) = (6-12) \times 10^{20}$ cm$^{-2}$.

\begin{figure*}
\centering
\vspace{5pt}
\includegraphics[scale=0.8]{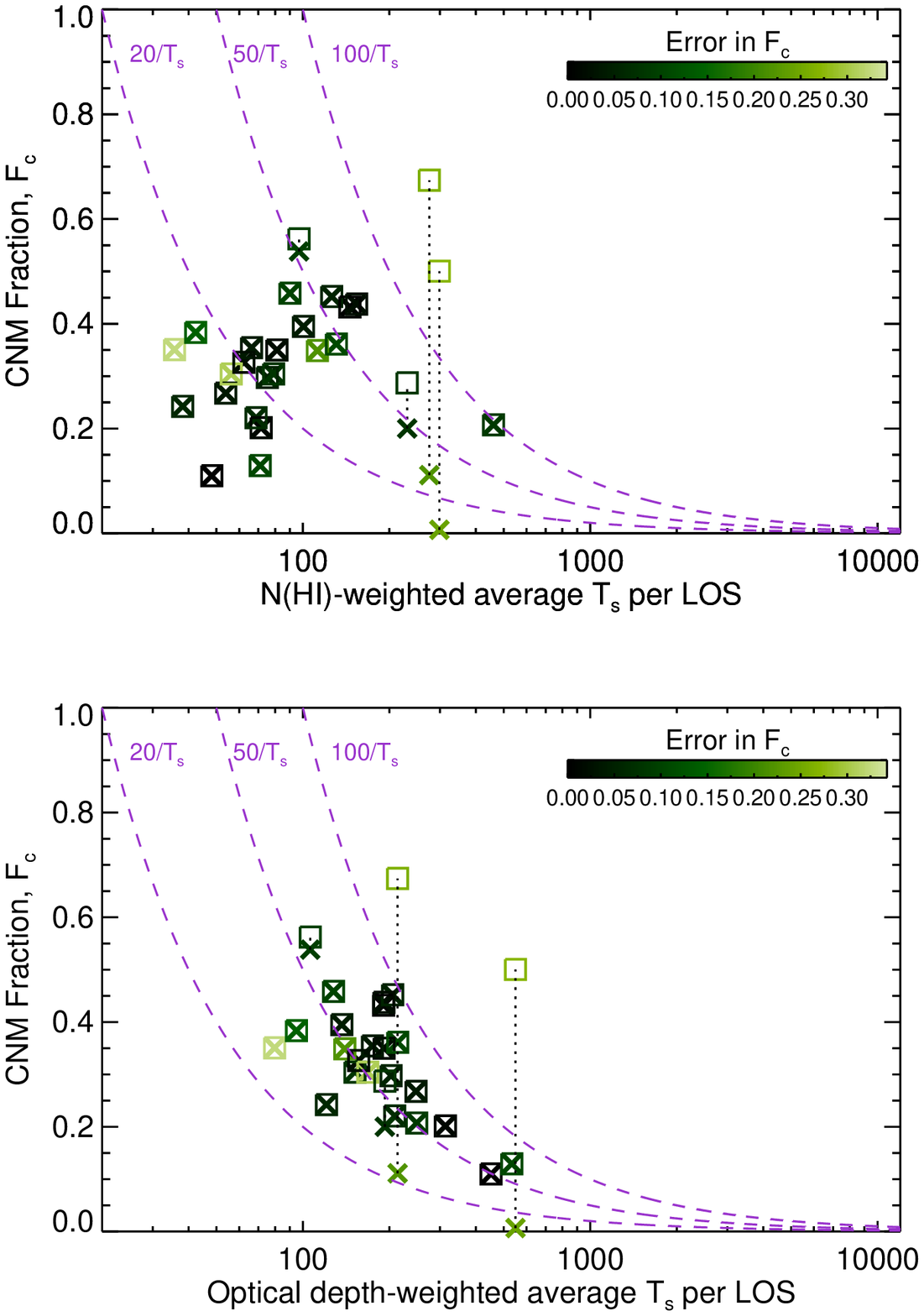}
\vspace{-75pt}
\caption{(Top) The CNM fraction as a function of the column density
weighted $T_s$ along each line of sight for our observations (crosses). 
Propagated uncertainties, shown as a color bar, are just from the fitting of Gaussian  
components and do not include any systematic uncertainties inherent to the
temperature derivation method. Crosses show the CNM fraction calculated
when components with $T_s<200$ K are considered as the CNM, while squares
show the $T_s<350$ K cut. The difference is very small and essentially 
only three sources have significantly changed their fraction.
(Bottom) The CNM fraction as a function of the optical-depth weighted
average $T_s$ (calculated using equation 15 in Kim et al. 2014), calculated
using the optical depth and $T_{exp}$ profiles without Gaussian fitting. 
Again, squares show that the temperature cut does not affect a majority of our sources.}
\label{f:dist_fract1}
\end{figure*}

In Figure~\ref{f:dist_fract1} (top) crosses show the CNM
fraction as a function of the column density weighted average spin temperature
along the line of sight.
A recent study by Kim et al. (2014), which
produced synthetic HI spectra based on their 3D hydrodynamic simulations of a Milky
Way-like disk, suggested that for the observed $T_s <400$ K the CNM fraction
is proportional to the inverse of  $T_s$.
While their synthetic spectra represent random directions,
most of the simulated data are located between $50 K/T_s$ and $100 K/T_s$ lines. 
Furthermore, for the observed $T_s <200$ K the simulated 
CNM fraction ranges between 40\% and 70\%, with a median value being 52\%
(97\% of simulated data points have a CNM fraction $<70$\%, Kim et al. 
private communication). 
We overplot the $1/T_s$ relation in the figure for three representative 
temperature values of 20, 50 and 100 K
(the simulation applied a CNM temperature 
cut-off of $T_k<184$ K, where $T_k$ is the true kinetic temperature ).
To bracket most of our data points we need to expand the
$T_s$ range to lower temperatures of $\la20$ K.
With our observed CNM fraction being largely in the range of 10-50\%, we 
overlap with the 40-70\% range expected by the simulation although,
the simulated fraction is generally slightly higher than what we observe.
Square symbols in this plot show the difference introduced in the CNM fraction 
when a 350 K threshold is applied (instead of 200 K) to select the CNM. 
The difference is very small, only three data points are noticeably affected.

In the same figure (bottom panel) instead of using our calculated $T_s$ we follow
exactly Kim et al. (2014) and calculate the observed temperature
as the optical-depth weighted average spin temperature along the line of sight
(equation (15) from Kim et al.). 
Most of our data points follow the 50/$T_s$ line, which agrees 
well with our median $T_s$ estimate, and is in 
excellent agreement with the Kim et al (2014) prediction.
Considering that in the simulation the CNM fraction is known, while
the observed CNM fraction is based on the Gaussian-decomposition 
$T_s$ derivation method, this excellent agreement is an indirect evidence that
the observational method provides consistent and reliable CNM fractions. 

While the optical-depth weighted average spin temperature (shown in the
bottom panel) is on average higher than our column density weighted spin temperature
(top panel), at the lowest temperatures the observed CNM fraction is
in the 10-50\% range, while the simulation suggests a CNM fraction of 40-70\%.
The simulated fraction is slightly higher that what is observed, however, it
is very encouraging to see that the simulated CNM fractions
are so close to observations and that the observed CNM fraction
follow the 50/$T_s$ prediction so closely.
Considering that Kim et al. (2014)
do not include interstellar chemistry, they likely slightly 
over-estimate the amount of cold HI
as the conversion from atomic to molecular phase is not taking place in the
simulation.

In summary, the CNM fraction in and close to Perseus is surprisingly low,
being largely below 50\% (median value of 30\%), 
even at the lowest observed temperature where HI absorption should
be tracing only the CNM with essentially no confusion by the thermally unstable WNM.
This is a somewhat surprising result as suggests that even close 
to the dense molecular clouds
the CNM fraction (CNM/CNM+WNM column density) is never very high.
As a consequence, this suggests that even lines of sight that probe deep
inside the GMCs have of order of 50\% contribution from the WNM 
(thermally unstable and/or stable). The geometry and the 
level of mixing of the CNM and WNM
are still not understood; for example it is not clear if the WNM
is located primarily in outer regions of the HI envelope, or is it being
brought closer to the inner regions via turbulence.
From the observational point of view, the HI absorption may not be tracing
the densest HI regions as optical depth profiles may become saturated,
like in the case of 4C+32.14 which is the source we had to exclude from analysis
due to its highly saturated HI absorption profile (this source is located
behind the main body of Perseus). 
It will be important to investigate the CNM fraction using alternative tracers in the future, 
like CII and CI.  

As the mixture of CNM and WNM phases exists in the diffuse ISM, \cite{Hennebelle06} asked the question 
of whether the WNM can persist deep inside molecular clouds. Considering that HI halos surround molecular clouds,
interstellar turbulence will naturally  mix in some HI with molecular gas.
However, in about one cooling time it is expected that any WNM mixed with molecular gas will
cool down if the internal pressure is about 10 times higher than the typical ISM pressure.
\cite{Hennebelle06} showed that the dissipation of magnetic waves can provide substantial
heating and therefore serve as an additional source of energy
that can maintain the WNM inside even high-pressure molecular clouds.

\section{Comparison of CO and HI absorption spectra}
\label{s:co_HI_abs}

\begin{figure*}
\centering
\includegraphics[scale=0.5]{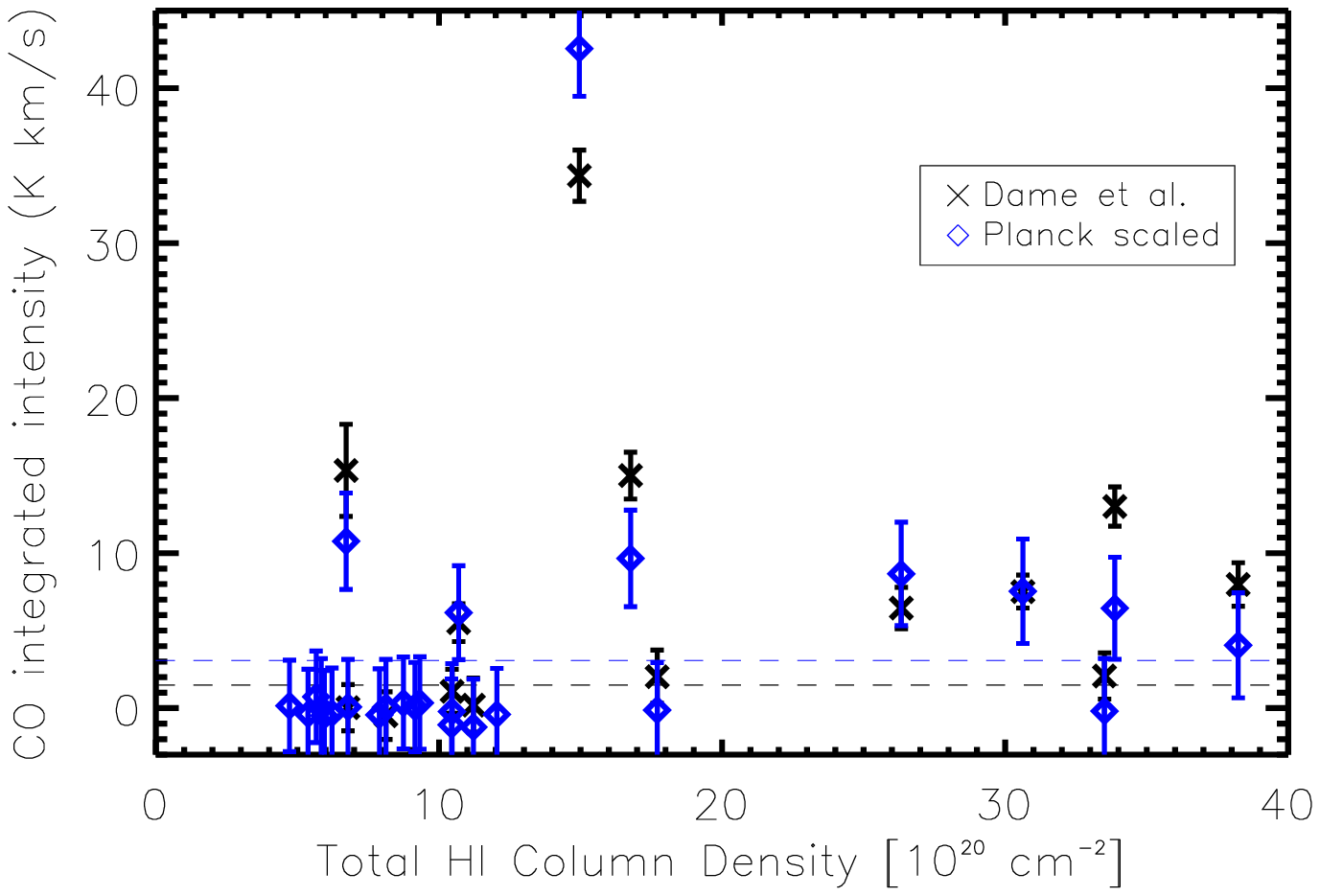}
\includegraphics[scale=0.5]{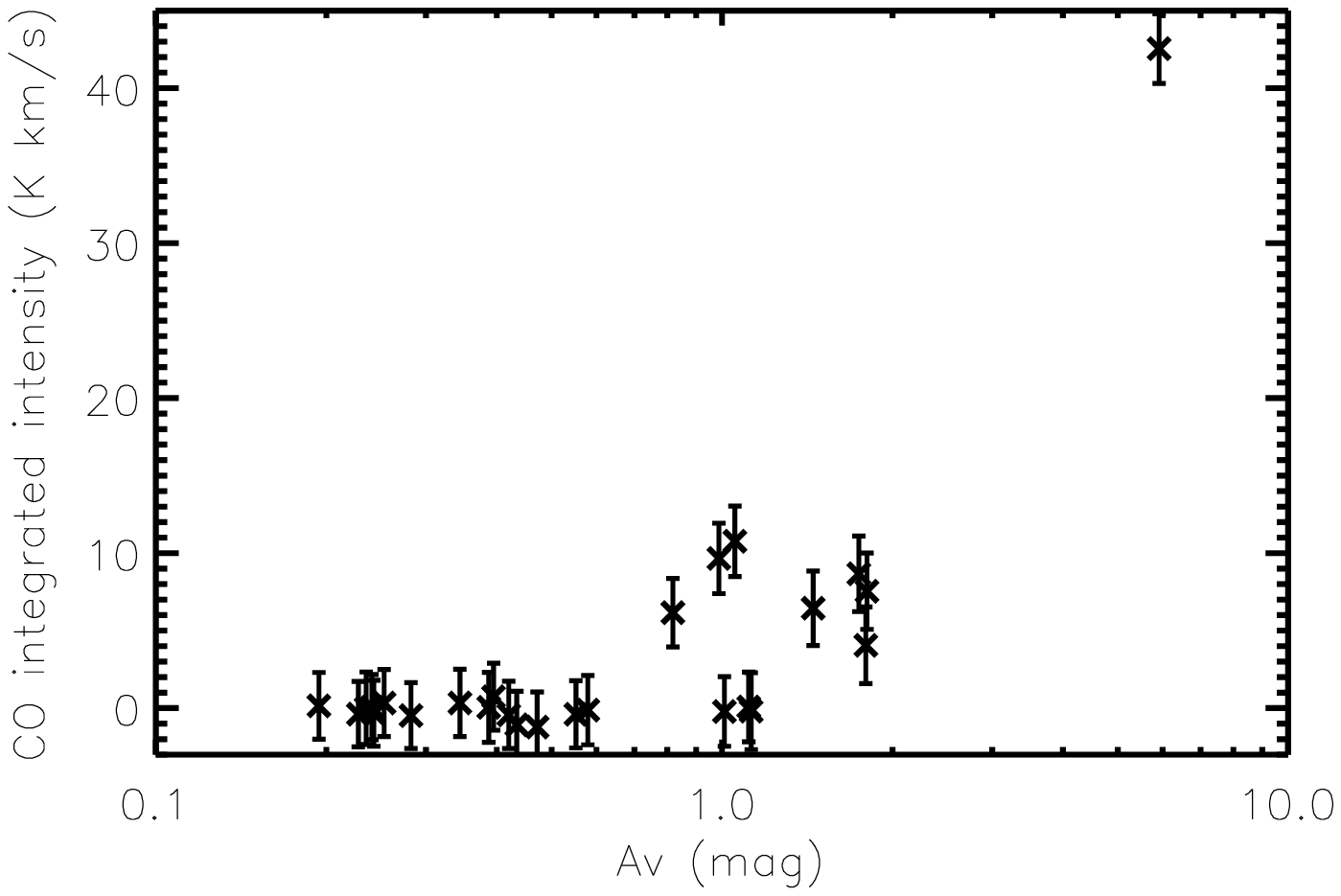}
\caption{(Left) CO integrated intensity as a function of the total HI column density
(CNM+WNM) in the direction of our 26 radio continuum sources.
CO data from Dame et al. (2001) are shown in black for directions covered in this survey; 
the black dashed line shows the median 1-$\sigma$ uncertainty.
CO data from Planck \citep{Planck13}  are shown in blue for all 26 sources after smoothing
and regridding  
the CO integrated image to 8.4$'$ resolution to match Dame et al.; the dashed blue lines shows
the median 1-$\sigma$ uncertainty. To match Planck and Dame et al. values a constant
scaling factor of 1.37 had to be applied to Planck data.
(Right) CO integrated intensity plotted as a function 
of $A_V$ from Planck.}
\label{f:Wco}
\end{figure*}

To compare HI absorption with CO we use data from two surveys:
the CO (1-0) emission data from the CfA survey at 8.4' resolution (Dame at al. 2001),
and the integrated CO intensity ($W_{CO}$) from Planck \citep{Planck13} 
smoothed and regridded to match the CfA's angular resolution and pixel size. 
We extract CO spectra from Dame et al. and show $W_{CO}$ in  
Figure~\ref{f:Wco} (left) as black data points. 
The Dame et al. observed area covers 14 out of 26 sources.
The dashed black line in this figure shows the median 1-$\sigma$ uncertainty
on $W_{CO}$ calculated from the line-free channels.
The results from Planck are shown in Figure~\ref{f:Wco} (left and right)
in blue, as well as their median 1-$\sigma$ uncertainty. 
We noticed that Dame et al.'s integrated intensity is systematically higher
relative to the Planck data. A median scaling of 1.37 was applied on the
Planck data to roughly match Dame et al.
observations. 

Figure~\ref{f:Wco} shows that 8 out of 26 sources have a clearly detected CO
emission that is above 1-$\sigma$ uncertainty in both Dame at al. and Planck data. 
Almost all detections have the total HI column density $>10^{21}$ cm$^{-2}$.
Their CNM fraction ranges from 20\% to 55\%.
While HI absorption is detected in the case of all sources,
18 sources were not detected in CO. 
Most non-detections pile up at $N(HI)<10^{21}$ cm$^{-2}$ and likely probe
diffuse HI regions.
As shown in Figure~\ref{f:Wco} (right) where we use the Planck
data for $E(B-V)\times 3.19$ as a measure of $A_V$ ($R_V=3.19$ was measured for 
Perseus star BD+31$^{\circ}$643 by \citep{Snow94}),
most non-detections have $A_V<1$.
Lee et al. (2014) showed that in Perseus $A_V\sim$1 mag 
is a necessary condition for the existence (shielding against photodissociation) 
of CO. Considering all this, most CO non-detections
probe diffuse ($A_V<1$ mag) regions without necessary shielding for CO formation.

However, there are 3 CO non-detections with $N(HI)=$10-35 
$\times 10^{20}$ cm$^{-2}$, which probe
regions with $A_V\sim1$ mag and therefore likely contain $H_2$, while CO could
be just forming and still be underabundant. 
Considering that we detect HI absorption with large column 
density, yet no CO emission, these three positions
are excellent candidates for probing the CO-dark gas which contains $H_2$ but not CO. 
Interestingly, CO is detected both at lower and higher total HI column density
relative to these non-detection,
at $\la10^{21}$ and $>3 \times 10^{21}$ cm$^{-2}$.
The three sources are: 3C132, 3C093.1, and B20411+34.
As shown in Figure 2, their HI absorption spectra have only components around 0 km/s
suggesting that a contamination from non-Perseus HI clouds can not be the reason
for HI absorption detections without CO emission.

\begin{figure*}
\centering
\vspace{-10pt}
\includegraphics[scale=0.75]{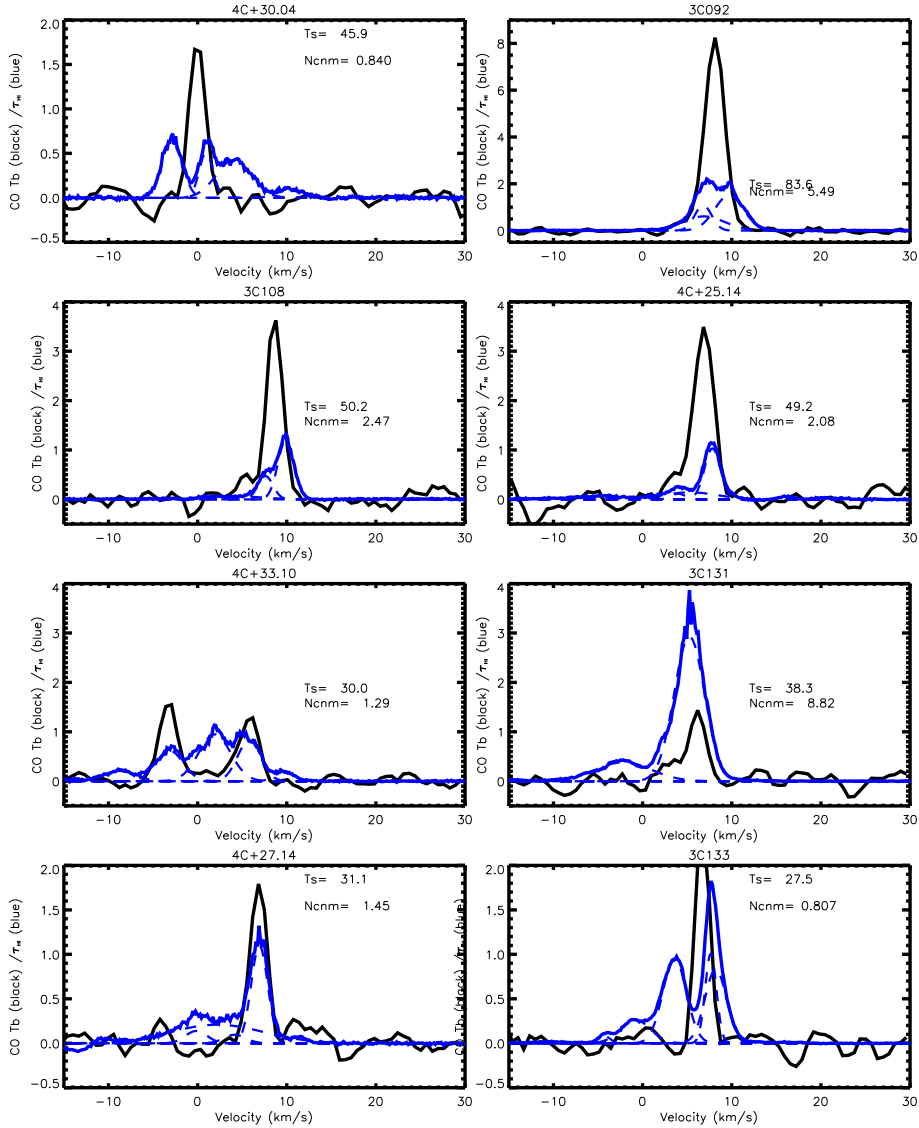}
\caption{Comparison of CO emission (from Dame et al.) shown in black and HI absorption shown in blue.
Each HI absorption profile has several Gaussian components shown as dashed lines.
For each panel, properties of the HI absorption component that is closest in
velocity to CO emission are listed in the top right corner: the corresponding spin temperature in K,
the corresponding CNM HI column density in $10^{20}$ cm$^{-2}$, and the CNM fraction calculated
for the whole line of sight.}
\label{f:co_HI_abs}
\end{figure*}

We now compare closely the kinematics of CO (from Dame et al.) 
and HI absorption of 8 sources with detected CO 
which have $A_V \ga 1$ mag (Figure~\ref{f:co_HI_abs}).
One of the eight sources, 3C092, is particularly interesting as 
it is located right behind the main body of Perseus, this source has the highest
integrated CO intensity ($>30$ K km s$^{-1}$) and the CNM fraction of $\sim0.4$. 
As shown in the figure, in most cases CO and HI absorption agree well in terms of
velocity range and profile shapes, although there is a large diversity among sources.
This suggests that
HI in absorption appears to trace not just cloud envelopes but also
central regions. In three cases (3C092, 3C108 and 4C+25.14)
CO emission and HI absorption cover the same velocity range.
In the case of 3C131, 3C133 and 4C+27.14, while the strongest HI absorption agrees well
with the CO emission peak, a weaker secondary 
component is seen at a velocity of 0 km/s which is not
detected in CO, possibly due to low sensitivity.
Only for two sources, 4C+30.04 and 4C+33.10 there is significant difference
in that the HI absorption profile is broader than CO emission and a CO peak is
found in the middle of the HI absorption profile.

We show in Figure~\ref{f:co_HI_abs} the corresponding spin temperature and 
the CNM column density of the HI component that is the closest 
in velocity to the CO peak. The spin temperature ranges from 30 to 80 K,
and the CNM column density of the component closest to CO 
ranges from 0.8 to 8 $\times 10^{20}$ cm$^{-2}$, 
which corresponds to the higher portion of 
the CNM column density measured for the whole population of CNM components in this study.
On the other hand, the remaining CNM column density along these lines of sight ranges from 
$\sim 3 \times 10^{20}$ to $13 \times 10^{20}$ cm$^{-2}$. 
All sources except 4C+25+14 have the total HI column density  $>10^{21}$ cm$^{-2}$ 
($A_V \ga 1$ mag), suggesting conditions suitable for formation of CO (and H$_2$).

This generally good spectral agreement we find between HI absorption and CO emission 
contrasts results from studies of the diffuse molecular
gas ($A_V<1$ mag), e.g. \cite{Liszt96,Liszt12}, where commonly HI absorption is more extended
in velocity relative to CO emission, and especially it was noticed that
CO emission tends to avoid the deepest HI absorption (in other words, CO was associated only with
weaker HI absorption features). This is usually explained as the deepest HI absorption arising 
mainly from the CO-free cloud envelopes, while CO tracing the central regions. 
The eight directions we investigate here all trace regions with $A_V\ga1$ and are
therefore likely probing equilibrium chemistry relative to $A_V<1$ likely largely
non-equilibrium dominated regions. 

It is generally expected that the CO-dark gas is found in uniform envelopes
surrounding CO-bright molecular clouds \citep{Wolfire10}.
Numerical simulations by \cite{Smith14} support this idea but show
that CO-dark H$_2$ may be asymmetric and not necessarily trace the
outlines of CO-bright clouds.
\cite{Fukui14a} proposed that the CO-dark
gas could be dominated by the optically thick HI.
In addition, considering that envelopes are likely 
to have small velocity offsets relative to the CO-bright cloud regions, 
we would expect to see kinematically more
extended HI absorption profiles around CO peaks.
However, in eight directions where we have both HI absorption and CO emission
spectra we generally find good agreement between the two.
This suggests that in these directions HI absorption traces largely the central 
cloud regions where CO is bright, and to a smaller degree
only the CO-dark cloud envelope. Of 26 directions there are only 3 cases
with strong HI absorption and the total $N(HI)>10^{21}$ cm$^{-2}$, but without CO emission.

Another interesting result from our study is that
cold HI with high HI column density is clearly present deep inside CO-bright GMCs, suggesting that its importance
for GMC evolution, and star formation, may be more significant than previously thought.
The origin of cold HI deep inside GMCs, and its morphology (e.g. filamentary
flows vs clumps vs diffuse distribution throughout GMCs)  are not well understood.
The cold HI could be brought deep into the clouds via circulation
of neutral gas from outer regions due to turbulence
\citep{Hennebelle06}, or could be a photodissociation product of H$_2$.
Tighter grids of HI absorption sources across and around GMCs are greatly needed 
to map out the distribution of cold HI and distinguish between various formation mechanisms.

\section{Summary  and Future Work}

To investigate properties of cold HI in and around Perseus molecular cloud, and
especially to investigate the role of cold HI in the shielding of H$_2$ (Paper II),
we have obtained and detected HI absorption in the direction of 26 background
radio continuum sources.
Using the corresponding HI emission spectra, and 
by employing a Gaussian decomposition of HI emission/absorption pairs, we have
performed radiative transfer calculations to estimate $T_s$ and $\tau(v)$ for
107 individual Gaussian components. 
This method represents the most direct way of measuring spin temperature and optical depth. 

The peak optical depth of individual Gaussian components ranges from $\sim0.01$ to a few, 
with the median value of 0.16. The spin temperature ranges from 10 to 725 K, and peaks at 50 K.
The median values of the CNM and WNM column densities for individual components are 
$6\times10^{19}$ cm$^{-2}$ and $1.5 \times 10^{20}$ cm$^{-2}$, respectively.
All properties of individual components for Perseus are in 
excellent agreement with those of HT03, who observed 
66 random lines of sight at $|b|>10^{\circ}$.
This suggests that
individual cold HI components have similar properties 
between  a focused field around a GMC and an average ISM field.

However, when all CNM and WNM components are summed 
along each line of sight, we find a significant
difference relative to an average ISM field. 
The Perseus region has a higher fraction of absorbing HI and a higher total HI 
column density relative to an average ISM field, suggesting environmental
differences.  This result is the first observational evidence that the CNM fraction 
in/around GMCs is likely higher than what is found in an average ISM field.
Interestingly, the median CNM and WNM HI column density along the line of sight
are roughly similar around Perseus,  
$4.6 \times 10^{20}$ cm$^{-2}$ vs $5.8 \times 10^{20}$ cm$^{-2}$, 
while in the case of HT03 the WNM column density was twice higher than the CNM column density.

Our results for both the optical depth and spin temperature are in stark contrast to
Fukui et al. (2014) who used Planck data and assumed that {\it all} of dust grains cooler than 22 K 
are mixed with the optically thick HI, suggesting that
the amount of cold HI could be significant and even enough to explain all (or
most) of the CO-dark gas. For 85\% of their sky coverage they estimated 
$T_s=20-40$ K and $\tau_{max}>0.5$. Considering all our Gaussian components,
we find such high $\tau_{max}$ only occasionally, with only 20\% of components
having $\tau_{max}>0.5$. Considering whole optical depth profiles, 54\% of directions have
$\tau_{max}>0.5$. 
Also, only $\sim15$\% of lines of sight have a column density weighted average
spin temperature lower than 40 K. 
We suspect that Fukui et al. results are caused by
the non ability to distinguish different gas components along the line of sight, as well
as by assigning {\it all} of the cooler dust to HI without allowing for
contribution of the molecular gas (bright or dark).

The mean spin temperature appears uniform over the radius of 10 to 120 pc
from the rough center of Perseus. Obtaining a tighter grid of HI absorption sources,
and especially sampling better the inner 10 pc, in the future will be important to probe a potential
radial increase in $T_s$ away from the cloud center as suggested by Andersson et al.
(1992).

While the CNM fraction is on average higher around Perseus 
relative to a random ISM field, surprisingly it rarely exceeds 50\%.
Even directions with the lowest $T_s<200$ K 
clearly show the CNM fraction of $<50$\%.
It is highly encouraging to see that recent numerical simulations by Kim et al.
(2014) produce the CNM fractions reasonably close to observations, $40-70$\%,
and also predict that the CNM fraction is inversely proportional to 
the optical-depth weighted
average $T_s$, which is in excellent agreement with observations.
Further inclusion of interstellar chemistry, and the HI-to-H$_2$ conversion, 
will likely fine-tune the simulated fractions and bring them even closer
to observations.
Our results suggest that even directions that probe deep inside molecular
clouds do not have high CNM fractions (e.g. $>50$\%). 
This could result from extended WNM envelopes
of GMCs, and/or significant mixing of CNM and WNM throughout GMCs caused by 
interstellar turbulence or accretion flows.
While the low CNM fraction in/around GMCs  requires further theoretical work, 
at high column densities the HI lines are likely to become saturated and 
therefore poorly trace the densest and coldest
regions of GMCs. It is therefore also important to observationally
test usefulness of additional tracers of neutral gas inside GMCs, e.g. CI and CII.

Finally, we have compared HI absorption with CO emission for our 26 directions and found
that 8/26 have detected CO. Out of the remaining 18, 15  directions
probe diffuse regions with $A_V<1$ mag and likely do not have enough shielding for CO formation.
Only 3/26 directions have $N(HI)>10^{21}$ cm$^{-2}$ ($A_V\ga1$ mag), and therefore
probe conditions suitable for CO formation, yet have no detected
CO emission. These directions therefore likely contain molecular gas but not CO
and are representative of so called CO-dark gas.
Eight directions with detected CO have $N(HI)>10^{21}$ cm$^{-2}$, $A_V>1$ mag,
and good kinematic agreement between HI absorption and CO emission spectra. All of this suggests
that these lines of sight probe largely central CO-bright regions, confirming the
existence of cold HI deep inside GMCs.
However, future observations of a tighter grid of background sources are necessary
to map out the distribution of cold HI around GMCs and its origin.

\begin{acknowledgements}
We sincerely thank telescope operators at the Arecibo Observatory for
their help in conducting these observations. The
Arecibo Observatory is operated by SRI International
under a cooperative agreement with the National Science
Foundation (AST-1100968), and in alliance with Ana G.
M\'endez-Universidad Metropolitana, and the Universities
Space Research Association.
We are extremely grateful to Chang-Goo Kim and Eve Ostriker
for extensive discussions and for providing detailed 
CNM fractions from Kim et al. (2014) for comparison
with observations.
We acknowledge stimulating discussions with Robert Lindner and Brian Babler, 
and thank Elijah Bernstein-Cooper for extracting Planck images around Perseus.
We also thank an anonymous referee for emphasizing the importance
of detailed chemistry in neutral gas estimates.
We are grateful to John Dickey for stressing the  importance of 
background diffuse emission in spin temperature calculations.
S.S. thanks the Department of Astronomy at the Faculty of 
Mathematics, Belgrade University
for their kind hospitality during the final stage of manuscript preparation.
This work was supported by the NSF Early Career
Development (CAREER) Award AST-1056780.
M.-Y. Lee acknowledges support from the DIM ACAV of the Region Ile de France.
We also acknowledge the NSF REU grant AST-1004881 which funded 
summer research of Jesse Miller. The use of ``Karma'' visualization
software (Gooch 1996) is gratefully acknowledged. 
\end{acknowledgements}


\centering
\label{tab:obs}
\tabcolsep=0.11cm
\begin{longtable}{lcccccccc}
\hline
\hline
\footnotesize
Source  & T$_B$ & V$_{\mathrm{LSR}}$ & $\Delta$V   & $\tau$ & T$_s$ & T$_{\mathrm{k,max}}$ & N(HI)$_{20}$ & $F$ or $O$  \\
        & (K)   & (km s$^{-1}$)     & (km s$^{-1}$)&        & (K)   & (K)                  & (cm$^{-2}$)  &          \\
        &  (1)      &  (2)        &  (3) 	  &    (4)    & (5)   &  (6)     & (7) & (8)  \\
\hline
\endhead
3C067
 &       1.88
 $\pm$       0.06 &      -25.8 $\pm$        0.1 &       4.71 $\pm$       0.18 & 
    0.0057 &       331. &          835 &       0.17 &        1.0 \\
3C067
 &       1.95
 $\pm$       0.00 &      -11.6 $\pm$        0.4 &      41.30 $\pm$       0.64 & 
    0.0045 &       438. &         2077 &       1.27 &        0.0 \\
3C067
 &       5.34
 &       -5.8 $\pm$        0.1 &       4.81 $\pm$       0.18 &       0.10 $\pm$ 
     0.002 &      69.10 $\pm$       8.88 &          325 &       0.62 & 
           2 \\
3C067
 &      21.02
 $\pm$       0.35 &       -2.3 $\pm$        0.2 &       9.75 $\pm$       0.17 & 
    0.0169 &      1254. &          861 &       3.99 &        0.5 \\
3C067
 &      23.05
 &       -0.3 $\pm$        0.0 &       2.14 $\pm$       0.04 &       0.41 $\pm$ 
     0.008 &      42.88 $\pm$       8.28 &          607 &       0.74 & 
           0 \\
3C067
 &      17.71
 &        1.2 $\pm$        0.1 &       5.99 $\pm$       0.13 &       0.19 $\pm$ 
     0.005 &      89.34 $\pm$      14.82 &          489 &       1.94 & 
           1 \\
3C067
 &       3.41
 $\pm$       0.19 &        8.3 $\pm$        0.2 &       6.28 $\pm$       0.31 & 
    0.0005 &      6748. &          485 &       0.41 &        0.0 \\
\hline
3C068.2
 &       1.24
 $\pm$       0.04 &      -19.8 $\pm$        1.4 &      30.94 $\pm$       2.61 & 
    0.0089 &       141. &        20919 &       0.55 &        0.0 \\
3C068.2
 &       5.09
 &       -6.2 $\pm$        0.0 &       2.58 $\pm$       0.07 &       0.27 $\pm$ 
     0.000 &      19.75 $\pm$       8.88 &          144 &       0.27 & 
           0 \\
3C068.2
 &      -0.09
 &       -3.0 $\pm$        0.1 &       2.63 $\pm$       0.24 &       0.09 $\pm$ 
     0.004 &       4.11 $\pm$       8.28 &          151 &       0.02 & 
           3 \\
3C068.2
 &      20.81
 $\pm$       1.16 &       -2.7 $\pm$        0.1 &      10.42 $\pm$       0.25 & 
    0.0147 &      1422. &         2370 &       4.22 &        0.0 \\
3C068.2
 &      31.03
 &        1.3 $\pm$        0.0 &       2.04 $\pm$       0.05 &       0.90 $\pm$ 
     0.040 &      43.82 $\pm$      14.82 &           91 &       1.57 & 
           2 \\
3C068.2
 &      16.32
 &        3.1 $\pm$        0.6 &       3.88 $\pm$       0.62 &       0.11 $\pm$ 
     0.017 &     140.94 $\pm$       5.04 &          328 &       1.15 & 
           1 \\
3C068.2
 &       3.44
 $\pm$       0.67 &        5.6 $\pm$        2.1 &      15.15 $\pm$       1.97 & 
    0.0194 &       179. &         5017 &       0.96 &        1.0 \\
\hline
3C092
 &       1.01
 $\pm$       0.06 &      -53.1 $\pm$        0.3 &       8.43 $\pm$       0.67 & 
    0.0117 &        86. &       119698 &       0.12 &        0.0 \\
3C092
 &       1.32
 $\pm$       0.04 &      -18.6 $\pm$        0.9 &      74.01 $\pm$       2.51 & 
    0.0059 &       224. &         3173 &       1.18 &        0.0 \\
3C092
 &       3.45
 $\pm$       0.06 &       -4.9 $\pm$        0.2 &      12.05 $\pm$       0.48 & 
    0.0031 &      1118. &         1661 &       0.81 &        0.0 \\
3C092
 &       6.36
 &        6.8 $\pm$        0.0 &       2.22 $\pm$       0.06 &       1.08 $\pm$ 
     0.040 &      26.89 $\pm$       8.28 &          151 &       1.26 & 
           2 \\
3C092
 &       6.36
 &        6.9 $\pm$        0.5 &       5.56 $\pm$       0.35 &       0.61 $\pm$ 
     0.114 &      49.09 $\pm$       8.88 &          144 &       3.23 & 
           1 \\
3C092
 &      49.81
 $\pm$       0.62 &        7.3 $\pm$        0.0 &       8.72 $\pm$       0.03 & 
    0.0025 &     19811. &          328 &       8.46 &        1.0 \\
3C092
 &       4.39
 &        9.7 $\pm$        0.1 &       3.82 $\pm$       0.12 &       1.50 $\pm$ 
     0.158 &      23.70 $\pm$      14.82 &           91 &       2.65 & 
           0 \\
\hline
3C093.1
 &       1.86
 $\pm$       0.10 &      -21.2 $\pm$        0.2 &       9.89 $\pm$       0.69 & 
    0.0048 &       386. &         1240 &       0.36 &        0.0 \\
3C093.1
 &       4.37
 $\pm$       0.07 &      -13.1 $\pm$        0.3 &      47.01 $\pm$       0.74 & 
    0.0016 &      2702. &          328 &       3.60 &        0.0 \\
3C093.1
 &      36.62
 &        6.2 $\pm$        0.0 &       1.62 $\pm$       0.09 &       0.41 $\pm$ 
     0.031 &      44.35 $\pm$       8.28 &          151 &       0.57 & 
           0 \\
3C093.1
 &      33.75
 $\pm$       0.59 &        7.2 $\pm$        0.0 &       7.53 $\pm$       0.06 & 
    0.0031 &     11047. &        48303 &       4.95 &        0.0 \\
3C093.1
 &      35.18
 &        7.7 $\pm$        0.0 &       4.45 $\pm$       0.04 &       1.19 $\pm$ 
     0.043 &      45.36 $\pm$       8.88 &          144 &       4.67 & 
           2 \\
3C093.1 &      35.58 &        8.8 $\pm$        0.0 &       1.36 $\pm$       0.03 &       1.30 $\pm$ 
     0.036 &      23.35 $\pm$      14.82 &           91 &       0.80 &           1 \\
\hline
3C108 &       1.91 $\pm$       0.05 &      -24.1 $\pm$        0.9 &      29.33 $\pm$       1.70 & 
    0.0045 &       425. &        18806 &       0.92 &        0.0 \\
3C108
 &       1.55
 &       -5.7 $\pm$        0.5 &       4.73 $\pm$       1.19 &       0.01 $\pm$ 
     0.002 &      64.12 $\pm$       8.88 &          144 &       0.07 & 
           2 \\
3C108
 &       5.21
 $\pm$       0.18 &       -0.7 $\pm$        0.1 &       4.69 $\pm$       0.14 & 
    0.0066 &       792. &          481 &       0.48 &        1.0 \\
3C108
 &      12.88
 &        4.0 $\pm$        0.3 &       7.52 $\pm$       0.69 &       0.06 $\pm$ 
     0.002 &     194.98 $\pm$       8.28 &          151 &       1.82 & 
           1 \\
3C108
 &       9.76
 $\pm$       0.19 &        6.7 $\pm$        0.3 &      23.59 $\pm$       0.31 & 
    0.0007 &     13061. &        12158 &       4.42 &        0.0 \\
3C108
 &      29.59
 &        7.6 $\pm$        0.0 &       1.99 $\pm$       0.05 &       0.47 $\pm$ 
     0.009 &      58.26 $\pm$      14.82 &           91 &       1.07 & 
           0 \\
3C108
 &      37.55
 &        9.9 $\pm$        0.0 &       2.10 $\pm$       0.02 &       1.20 $\pm$ 
     0.010 &      49.52 $\pm$       5.04 &          328 &       2.44 & 
           3 \\
\hline
3C131
 &      16.27
 &      -39.5 $\pm$        0.0 &       3.74 $\pm$       0.09 &       0.16 $\pm$ 
     0.003 &     100.90 $\pm$       8.88 &          144 &       1.15 &       2
 \\
3C131
 &       3.41
 &      -34.0 $\pm$        0.1 &       2.00 $\pm$       0.33 &       0.03 $\pm$ 
     0.004 &      75.15 $\pm$       8.28 &          151 &       0.09 &       3
 \\
3C131
 &       2.28
 &      -24.1 $\pm$        0.2 &       3.37 $\pm$       0.50 &       0.03 $\pm$ 
     0.003 &     114.31 $\pm$      14.82 &           91 &       0.20 &       4
 \\
3C131
 &       9.18
 $\pm$       0.15 &      -18.5 $\pm$        0.3 &      53.41 $\pm$       0.68 & 
    0.0079 &      1163. &          745 &       8.82 &        0.0 \\
3C131
 &       4.62
 &      -15.9 $\pm$        0.3 &       6.76 $\pm$       0.70 &       0.03 $\pm$ 
     0.002 &     227.36 $\pm$       5.04 &          328 &       0.89 &       1
 \\
3C131
 &       7.18
 $\pm$       0.26 &       -9.6 $\pm$        0.1 &       4.94 $\pm$       0.21 & 
    0.0036 &      2023. &          745 &       0.69 &        0.0 \\
3C131
 &      46.95
 &       -2.1 $\pm$        0.0 &       7.31 $\pm$       0.10 &       0.38 $\pm$ 
     0.003 &     145.62 $\pm$       6.75 &        62335 &       7.99 &       0
 \\
3C131
 &      46.76
 &        5.2 $\pm$        0.0 &       4.04 $\pm$       0.02 &       2.93 $\pm$ 
     0.028 &      37.71 $\pm$       6.26 &          534 &       8.68 &       5
 \\
3C131
 &      45.98
 $\pm$       1.89 &        5.9 $\pm$        0.0 &       5.84 $\pm$       0.10 & 
    0.0234 &      1985. &          745 &       5.23 &        0.0 \\
3C131
 &       6.36
 &       11.9 $\pm$        0.2 &       2.65 $\pm$       0.54 &       0.02 $\pm$ 
     0.004 &     119.41 $\pm$       6.19 &          745 &       0.13 &       6
 \\
\hline
3C132
 &       6.16
 $\pm$       0.03 &      -15.4 $\pm$        0.2 &      50.11 $\pm$       0.37 & 
    0.0164 &       378. &        54879 &       5.68 &        0.0 \\
3C132
 &       2.21
 &       -2.3 $\pm$        0.1 &       3.15 $\pm$       0.28 &       0.05 $\pm$ 
     0.003 &      37.05 $\pm$       8.88 &          144 &       0.12 & 
           0 \\
3C132
 &      11.81
 &        2.0 $\pm$        0.0 &       4.12 $\pm$       0.07 &       0.34 $\pm$ 
     0.003 &      47.85 $\pm$       8.28 &          151 &       1.29 & 
           2 \\
3C132
 &      38.49
 $\pm$       0.19 &        5.3 $\pm$        0.0 &      14.52 $\pm$       0.05 & 
    0.0068 &      5703. &         4610 &      10.89 &        0.5 \\
3C132
 &      12.35
 &        8.0 $\pm$        0.0 &       2.54 $\pm$       0.01 &       1.33 $\pm$ 
     0.008 &      35.41 $\pm$      14.82 &           91 &       2.32 & 
           3 \\
3C132
 &      54.35
 $\pm$       2.87 &        8.4 $\pm$        0.0 &       3.18 $\pm$       0.02 & 
    0.0015 &     35172. &          220 &       3.36 &        0.5 \\
3C132
 &      23.31
 &       13.1 $\pm$        0.0 &       5.24 $\pm$       0.08 &       0.24 $\pm$ 
     0.002 &     109.33 $\pm$       5.04 &          328 &       2.66 & 
           1 \\
\hline
3C133
 &      15.84
 &      -29.8 $\pm$        0.1 &       8.18 $\pm$       0.17 &       0.05 $\pm$ 
     0.002 &     322.42 $\pm$       8.88 &         1462 &       2.52 & 
           0 \\
3C133
 &      14.58
 &      -27.7 $\pm$        0.0 &       3.28 $\pm$       0.11 &       0.07 $\pm$ 
     0.002 &      38.56 $\pm$       8.28 &          234 &       0.17 & 
           3 \\
3C133
 &      10.16
 $\pm$       0.08 &      -16.8 $\pm$        0.3 &      43.40 $\pm$       0.44 & 
    0.0012 &      8199. &        41156 &       8.26 &        0.5 \\
3C133
 &       3.20
 &      -10.3 $\pm$        0.1 &       5.98 $\pm$       0.38 &       0.02 $\pm$ 
     0.001 &     219.63 $\pm$      14.82 &          782 &       0.40 & 
           5 \\
3C133
 &       0.77
 &       -4.3 $\pm$        0.0 &       1.49 $\pm$       0.05 &       0.08 $\pm$ 
     0.002 &       4.89 $\pm$       5.04 &           48 &       0.01 & 
           6 \\
3C133
 &       5.98
 &       -0.8 $\pm$        0.0 &       3.96 $\pm$       0.07 &       0.26 $\pm$ 
     0.001 &      12.37 $\pm$       6.75 &          342 &       0.25 & 
           7 \\
3C133
 &      67.68
 $\pm$       0.54 &        2.5 $\pm$        0.0 &      10.40 $\pm$       0.05 & 
    0.0009 &     73191. &         2362 &      13.71 &        0.0 \\
3C133
 &      42.16
 &        3.6 $\pm$        0.0 &       3.10 $\pm$       0.02 &       0.95 $\pm$ 
     0.003 &      47.99 $\pm$       6.26 &          210 &       2.77 & 
           8 \\
3C133
 &      48.15
 &        7.6 $\pm$        0.0 &       1.49 $\pm$       0.02 &       1.01 $\pm$ 
     0.019 &      10.82 $\pm$      36.16 &           48 &       0.32 & 
           2 \\
3C133
 &      49.11
 &        8.2 $\pm$        0.0 &       2.84 $\pm$       0.02 &       0.83 $\pm$ 
     0.017 &      69.32 $\pm$       6.19 &          176 &       3.19 & 
           4 \\
3C133
 &      11.35
 &       11.9 $\pm$        1.1 &       8.75 $\pm$       1.56 &       0.02 $\pm$ 
     0.001 &     725.42 $\pm$      17.67 &         1675 &       1.91 & 
           1 \\
\hline
4C+25.14
 &       1.86
 $\pm$       0.06 &      -23.5 $\pm$        1.1 &      30.47 $\pm$       1.82 & 
    0.0147 &       127. &        20291 &       0.95 &        0.5 \\
4C+25.14
 &       0.14
 &       -8.3 $\pm$        4.5 &      16.30 $\pm$       5.51 &       0.02 $\pm$ 
     0.006 &      12.84 $\pm$       8.88 &         5802 &       0.09 & 
           2 \\
4C+25.14
 &       1.13
 &       -4.7 $\pm$        0.2 &       4.59 $\pm$       0.88 &       0.04 $\pm$ 
     0.014 &      21.83 $\pm$       8.28 &          462 &       0.09 & 
           0 \\
4C+25.14
 &      20.06
 &        4.0 $\pm$        0.1 &       1.82 $\pm$       0.18 &       0.11 $\pm$ 
     0.011 &      41.54 $\pm$      14.82 &           73 &       0.17 & 
           4 \\
4C+25.14
 &      19.63
 &        5.3 $\pm$        0.2 &       7.15 $\pm$       0.38 &       0.14 $\pm$ 
     0.016 &     130.63 $\pm$       5.04 &         1117 &       2.61 & 
           3 \\
4C+25.14
 &       8.95
 $\pm$       0.11 &        6.0 $\pm$        0.3 &      27.53 $\pm$       0.40 & 
    0.0022 &      4025. &        16562 &       4.69 &        0.0 \\
4C+25.14
 &      36.35
 &        7.8 $\pm$        0.0 &       2.10 $\pm$       0.03 &       1.03 $\pm$ 
     0.012 &      41.83 $\pm$       6.75 &           96 &       1.77 & 
           1 \\
4C+25.14
 &       1.17
 &       16.1 $\pm$        0.1 &       1.76 $\pm$       0.13 &       0.06 $\pm$ 
     0.004 &      15.94 $\pm$       6.26 &           67 &       0.03 & 
           5 \\
4C+25.14
 &       0.90
 &       20.9 $\pm$        0.2 &       3.82 $\pm$       0.41 &       0.03 $\pm$ 
     0.003 &      32.05 $\pm$       6.19 &          318 &       0.07 & 
           6 \\
\hline
4C+26.12
 &       7.28
 &       -7.0 $\pm$        0.0 &       2.60 $\pm$       0.05 &       0.16 $\pm$ 
     0.003 &      53.06 $\pm$       8.88 &         5802 &       0.44 & 
           1 \\
4C+26.12
 &       1.45
 $\pm$       0.04 &       -6.1 $\pm$        0.6 &      59.78 $\pm$       1.92 & 
    0.0049 &       295. &         1117 &       0.95 &        1.0 \\
4C+26.12
 &       6.59
 $\pm$       0.11 &       -3.9 $\pm$        0.2 &       7.30 $\pm$       0.26 & 
    0.0032 &      2088. &           96 &       0.94 &        1.0 \\
4C+26.12
 &       4.61
 $\pm$       0.29 &        1.7 $\pm$        0.1 &       4.46 $\pm$       0.24 & 
    0.0008 &      6111. &           67 &       0.40 &        0.0 \\
4C+26.12
 &      28.22
 &        6.6 $\pm$        0.1 &       8.95 $\pm$       0.20 &       0.11 $\pm$ 
     0.003 &     182.48 $\pm$      14.82 &           73 &       3.50 & 
           0 \\
4C+26.12
 &      28.39
 &        6.7 $\pm$        0.0 &       1.83 $\pm$       0.04 &       0.30 $\pm$ 
     0.005 &      53.04 $\pm$       8.28 &          462 &       0.56 & 
           2 \\
\hline
4C+27.07
 &       2.09
 $\pm$       0.02 &      -23.0 $\pm$        0.0 &      24.44 $\pm$       0.27 & 
    0.0124 &       169. &           73 &       0.86 &        1.0 \\
4C+27.07
 &       8.35
 &       -5.2 $\pm$        0.1 &       3.67 $\pm$       0.16 &       0.10 $\pm$ 
     0.003 &      87.61 $\pm$       8.88 &         5802 &       0.60 & 
           0 \\
4C+27.07
 &      16.85
 $\pm$       0.14 &       -1.7 $\pm$        0.0 &      10.73 $\pm$       0.05 & 
    0.0224 &       761. &           96 &       3.52 &        0.0 \\
4C+27.07
 &      15.70
 &       -0.3 $\pm$        0.0 &       2.96 $\pm$       0.05 &       0.27 $\pm$ 
     0.004 &      61.24 $\pm$       8.28 &          462 &       0.94 & 
           1 \\
4C+27.07
 &       2.24
 $\pm$       0.12 &        3.1 $\pm$        0.3 &      20.01 $\pm$       0.32 & 
    0.0023 &       971. &         1117 &       0.79 &        0.0 \\
\hline
4C+27.14
 &      -1.05
 &      -41.5 $\pm$        0.1 &       1.74 $\pm$       0.18 &       0.08 $\pm$ 
     0.007 &       3.58 $\pm$       8.88 &           66 &       0.01 &       4
 \\
4C+27.14
 &      13.09
 &      -35.2 $\pm$        0.0 &       2.49 $\pm$       0.11 &       0.29 $\pm$ 
     0.012 &      34.00 $\pm$       8.28 &          135 &       0.47 &       5
 \\
4C+27.14
 &      14.70
 $\pm$       0.24 &      -34.1 $\pm$        0.1 &      20.31 $\pm$       0.25 & 
    0.0172 &       864. &         9019 &       5.80 &        0.0 \\
4C+27.14
 &       8.28
 &      -31.2 $\pm$        0.3 &       7.65 $\pm$       0.43 &       0.14 $\pm$ 
     0.005 &      65.97 $\pm$       5.04 &         1278 &       1.33 &       7
 \\
4C+27.14
 &       6.28
 &      -28.9 $\pm$        0.0 &       1.19 $\pm$       0.07 &       0.23 $\pm$ 
     0.011 &      13.99 $\pm$      14.82 &           30 &       0.07 &       6
 \\
4C+27.14
 &       3.20
 &      -20.7 $\pm$        0.2 &       3.89 $\pm$       0.42 &       0.05 $\pm$ 
     0.004 &      85.24 $\pm$       6.75 &          330 &       0.32 &       8
 \\
4C+27.14
 &      12.75
 $\pm$       0.41 &      -12.9 $\pm$        0.1 &       8.42 $\pm$       0.27 & 
    0.0052 &      2478. &         1548 &       2.09 &        0.0 \\
4C+27.14
 &      34.04
 &       -0.1 $\pm$        0.1 &       2.84 $\pm$       0.31 &       0.14 $\pm$ 
     0.016 &      45.07 $\pm$       6.26 &          176 &       0.35 &       1
 \\
4C+27.14
 &      33.23
 &        2.2 $\pm$        0.1 &      12.26 $\pm$       0.46 &       0.21 $\pm$ 
     0.020 &     161.68 $\pm$      36.16 &         3287 &       8.06 &       0
 \\
4C+27.14
 &      19.03
 $\pm$       1.55 &        2.8 $\pm$        0.3 &      19.60 $\pm$       0.50 & 
    0.0145 &      1320. &         8396 &       7.23 &        0.0 \\
4C+27.14
 &      28.87
 &        4.2 $\pm$        0.3 &       2.86 $\pm$       0.74 &       0.10 $\pm$ 
     0.019 &      13.99 $\pm$       6.19 &          178 &       0.07 &       3
 \\
4C+27.14
 &      38.30
 $\pm$       2.39 &        5.0 $\pm$        0.0 &       5.66 $\pm$       0.14 & 
    0.0044 &      8820. &          700 &       4.23 &        0.0 \\
4C+27.14
 &      40.88
 &        6.9 $\pm$        0.0 &       2.18 $\pm$       0.05 &       1.10 $\pm$ 
     0.022 &      12.82 $\pm$      17.67 &          103 &       0.60 &       2
 \\
\hline
4C+28.06
 &       2.07
 $\pm$       0.05 &       -6.9 $\pm$        0.3 &      39.13 $\pm$       0.60 & 
    0.0021 &       984. &          330 &       1.23 &        1.0 \\
4C+28.06
 &       4.83
 &       -6.8 $\pm$        0.1 &       4.80 $\pm$       0.17 &       0.07 $\pm$ 
     0.002 &      73.35 $\pm$       8.88 &           66 &       0.51 & 
           2 \\
4C+28.06
 &      16.89
 $\pm$       0.12 &        0.3 $\pm$        0.0 &      11.72 $\pm$       0.11 & 
    0.0058 &      2937. &          176 &       3.85 &        1.0 \\
4C+28.06
 &      14.32
 &        1.5 $\pm$        0.0 &       5.31 $\pm$       0.07 &       0.28 $\pm$ 
     0.004 &      69.60 $\pm$       8.28 &          135 &       2.02 & 
           3 \\
4C+28.06
 &      12.98
 &        3.3 $\pm$        0.0 &       1.85 $\pm$       0.06 &       0.24 $\pm$ 
     0.007 &      27.27 $\pm$      14.82 &           30 &       0.24 & 
           1 \\
4C+28.06
 &       1.16
 &        7.8 $\pm$        0.1 &       1.70 $\pm$       0.23 &       0.03 $\pm$ 
     0.003 &      36.98 $\pm$       5.04 &         1278 &       0.04 & 
           0 \\
\hline
4C+28.07
 &       0.73
 $\pm$       0.02 &      -16.5 $\pm$        0.7 &      56.11 $\pm$       1.46 & 
    0.0034 &       212. &          178 &       0.35 &        0.0 \\
4C+28.07
 &      20.10
 &       -6.3 $\pm$        0.0 &       2.36 $\pm$       0.10 &       0.15 $\pm$ 
     0.009 &      58.15 $\pm$       8.28 &          135 &       0.40 & 
           1 \\
4C+28.07
 &      19.34
 &       -5.7 $\pm$        0.1 &       6.19 $\pm$       0.42 &       0.08 $\pm$ 
     0.009 &     169.27 $\pm$       8.88 &           66 &       1.71 & 
           2 \\
4C+28.07
 &      11.90
 &        0.7 $\pm$        0.1 &       3.02 $\pm$       0.08 &       0.57 $\pm$ 
     0.019 &      16.40 $\pm$      14.82 &           30 &       0.55 & 
           4 \\
4C+28.07
 &       9.04
 $\pm$       0.28 &        1.2 $\pm$        0.1 &      16.08 $\pm$       0.13 & 
    0.0019 &      4685. &          176 &       2.79 &        0.0 \\
4C+28.07
 &      31.30
 $\pm$       0.71 &        1.4 $\pm$        0.0 &       4.65 $\pm$       0.02 & 
    0.0045 &      6920. &         3287 &       2.83 &        0.0 \\
4C+28.07
 &       6.65
 &        3.2 $\pm$        0.2 &       2.94 $\pm$       0.25 &       0.19 $\pm$ 
     0.026 &      39.45 $\pm$       5.04 &         1278 &       0.42 & 
           0 \\
4C+28.07
 &       0.99
 &        6.7 $\pm$        1.4 &       6.25 $\pm$       1.93 &       0.02 $\pm$ 
     0.003 &     139.49 $\pm$       6.75 &          330 &       0.28 & 
           3 \\
\hline
4C+29.05
 &       1.57
 $\pm$       0.03 &      -11.4 $\pm$        0.2 &      43.60 $\pm$       0.47 & 
    0.0160 &        99. &          330 &       0.96 &        1.0 \\
4C+29.05
 &       0.67
 &       -8.0 $\pm$        0.1 &       2.20 $\pm$       0.31 &       0.03 $\pm$ 
     0.003 &      32.58 $\pm$       8.88 &           66 &       0.04 & 
           0 \\
4C+29.05
 &       6.13
 $\pm$       0.08 &       -4.4 $\pm$        0.1 &      16.62 $\pm$       0.13 & 
    0.0082 &       752. &         1278 &       1.97 &        1.0 \\
4C+29.05
 &       6.28
 &       -1.3 $\pm$        0.0 &       3.93 $\pm$       0.08 &       0.14 $\pm$ 
     0.003 &      49.20 $\pm$       8.28 &          135 &       0.54 & 
           1 \\
4C+29.05
 &       7.87
 $\pm$       0.10 &       -0.4 $\pm$        0.0 &       7.94 $\pm$       0.07 & 
    0.0017 &      4770. &           30 &       1.22 &        0.0 \\
\hline
4C+30.04
 &       1.30
 $\pm$       0.04 &       -4.8 $\pm$        0.7 &      42.10 $\pm$       1.04 & 
    0.0191 &        68. &        38738 &       0.74 &        1.0 \\
4C+30.04
 &      23.10
 &       -2.9 $\pm$        0.0 &       2.62 $\pm$       0.02 &       0.67 $\pm$ 
     0.007 &      45.83 $\pm$       8.88 &          149 &       1.57 & 
           1 \\
4C+30.04
 &      19.23
 &        1.0 $\pm$        0.0 &       1.83 $\pm$       0.05 &       0.51 $\pm$ 
     0.013 &      40.71 $\pm$       8.28 &           73 &       0.74 & 
           3 \\
4C+30.04
 &      35.95
 $\pm$       0.22 &        2.3 $\pm$        0.0 &      10.46 $\pm$       0.03 & 
    0.0109 &      3326. &         2391 &       7.33 &        0.0 \\
4C+30.04
 &       7.00
 &        4.1 $\pm$        0.1 &       4.61 $\pm$       0.12 &       0.43 $\pm$ 
     0.004 &      15.64 $\pm$      14.82 &          465 &       0.60 & 
           2 \\
4C+30.04
 &       0.09
 &       10.4 $\pm$        0.1 &       3.16 $\pm$       0.15 &       0.10 $\pm$ 
     0.004 &       1.99 $\pm$       5.04 &          217 &       0.01 & 
           0 \\
4C+30.04
 &       4.26
 $\pm$       0.11 &       13.7 $\pm$        0.3 &      13.23 $\pm$       0.50 & 
    0.0161 &       267. &         3822 &       1.06 &        0.0 \\
\hline
4C+33.10
 &       4.68
 &      -52.8 $\pm$        0.1 &       2.36 $\pm$       0.13 &       0.12 $\pm$ 
     0.006 &      33.86 $\pm$       8.88 &          149 &       0.19 & 
           4 \\
4C+33.10
 &      12.70
 $\pm$       0.39 &      -46.2 $\pm$        0.2 &      17.41 $\pm$       0.43 & 
    0.0037 &      3394. &         1017 &       4.29 &        0.0 \\
4C+33.10
 &      10.10
 &      -44.8 $\pm$        0.0 &       3.06 $\pm$       0.09 &       0.20 $\pm$ 
     0.006 &      52.19 $\pm$       8.28 &           73 &       0.62 & 
           5 \\
4C+33.10
 &      12.34
 $\pm$       1.36 &      -43.2 $\pm$        0.3 &       4.50 $\pm$       0.29 & 
    0.0039 &      3180. &         1017 &       1.08 &        0.0 \\
4C+33.10
 &       5.44
 $\pm$       0.30 &      -30.6 $\pm$        0.2 &       6.82 $\pm$       0.54 & 
    0.0026 &      2081. &         1017 &       0.72 &        0.0 \\
4C+33.10
 &       3.88
 &      -16.1 $\pm$        0.1 &       2.77 $\pm$       0.25 &       0.07 $\pm$ 
     0.005 &      45.43 $\pm$      14.82 &          465 &       0.17 & 
           6 \\
4C+33.10
 &       4.41
 &       -8.8 $\pm$        0.1 &       4.27 $\pm$       0.17 &       0.21 $\pm$ 
     0.005 &      19.31 $\pm$       5.04 &          217 &       0.33 & 
           7 \\
4C+33.10
 &      34.42
 $\pm$       0.45 &       -6.8 $\pm$        0.1 &      24.91 $\pm$       0.23 & 
    0.0118 &      2933. &         1017 &      16.62 &        0.0 \\
4C+33.10
 &      23.35
 &       -3.2 $\pm$        0.1 &       3.57 $\pm$       0.12 &       0.62 $\pm$ 
     0.010 &      48.13 $\pm$       6.75 &          443 &       2.07 & 
           0 \\
4C+33.10
 &      43.26
 &        1.9 $\pm$        0.1 &       4.21 $\pm$       0.21 &       0.95 $\pm$ 
     0.012 &      69.69 $\pm$       6.26 &        13556 &       5.43 & 
           1 \\
4C+33.10
 &      42.58
 &        5.8 $\pm$        0.1 &       3.03 $\pm$       0.11 &       0.83 $\pm$ 
     0.028 &      70.14 $\pm$       6.19 &          684 &       3.44 & 
           2 \\
4C+33.10
 &      26.72
 $\pm$       1.09 &        7.3 $\pm$        0.2 &       5.59 $\pm$       0.17 & 
    0.0116 &      2317. &         1017 &       2.91 &        0.0 \\
4C+33.10
 &       7.53
 &        9.7 $\pm$        0.1 &       2.46 $\pm$       0.14 &       0.19 $\pm$ 
     0.007 &      36.57 $\pm$      36.16 &         6621 &       0.34 & 
           3 \\
\hline
4C+34.07
 &       2.00
 $\pm$       0.00 &      -22.7 $\pm$        0.0 &       3.07 $\pm$       0.10 & 
    0.0132 &       153. &         3403 &       0.12 &        1.0 \\
4C+34.07
 &       1.70
 $\pm$       0.03 &       -9.1 $\pm$        0.3 &      46.64 $\pm$       0.63 & 
    0.0002 &     10602. &          217 &       1.14 &        0.0 \\
4C+34.07
 &      20.20
 &       -2.2 $\pm$        0.0 &       1.84 $\pm$       0.05 &       0.24 $\pm$ 
     0.007 &      21.05 $\pm$       8.88 &          149 &       0.18 & 
           0 \\
4C+34.07
 &      20.62
 &       -1.0 $\pm$        0.0 &       5.63 $\pm$       0.10 &       0.15 $\pm$ 
     0.009 &     140.20 $\pm$       8.28 &           73 &       2.26 & 
           2 \\
4C+34.07
 &       7.21
 $\pm$       0.09 &        0.3 $\pm$        0.0 &      12.48 $\pm$       0.11 & 
    0.0086 &       845. &        47549 &       1.75 &        0.0 \\
4C+34.07
 &      18.91
 &        1.1 $\pm$        0.0 &       1.79 $\pm$       0.11 &       0.09 $\pm$ 
     0.006 &      63.79 $\pm$      14.82 &          465 &       0.21 & 
           1 \\
\hline
4C+34.09
 &       2.30
 $\pm$       0.05 &      -10.2 $\pm$        0.5 &      61.68 $\pm$       1.46 & 
    0.0051 &       452. &          217 &       2.18 &        0.0 \\
4C+34.09
 &       0.74
 &       -8.0 $\pm$        0.0 &       1.69 $\pm$       0.03 &       0.22 $\pm$ 
     0.004 &       4.79 $\pm$       8.88 &          149 &       0.03 & 
           2 \\
4C+34.09
 &       7.65
 $\pm$       0.12 &       -2.1 $\pm$        0.1 &       8.36 $\pm$       0.20 & 
    0.0017 &      4607. &        83144 &       1.25 &        1.0 \\
4C+34.09
 &      35.09
 &        1.1 $\pm$        0.0 &       2.70 $\pm$       0.06 &       0.24 $\pm$ 
     0.005 &      93.71 $\pm$       8.28 &           73 &       1.19 & 
           0 \\
4C+34.09
 &      27.38
 &        4.7 $\pm$        0.1 &       9.80 $\pm$       0.11 &       0.19 $\pm$ 
     0.002 &     157.27 $\pm$      14.82 &          465 &       5.79 & 
           1 \\
\hline
5C06.237
 &       2.30
 $\pm$       0.05 &       -7.4 $\pm$        0.3 &      40.50 $\pm$       0.71 & 
    0.0031 &       750. &          217 &       1.48 &        0.0 \\
5C06.237
 &      14.03
 &       -1.3 $\pm$        0.0 &       3.42 $\pm$       0.07 &       0.40 $\pm$ 
     0.004 &      52.28 $\pm$       8.88 &          149 &       1.40 & 
           1 \\
5C06.237
 &      14.48
 $\pm$       0.20 &       -0.5 $\pm$        0.0 &       9.96 $\pm$       0.10 & 
    0.0094 &      1545. &          465 &       2.81 &        1.0 \\
5C06.237
 &       6.37
 &        2.0 $\pm$        0.1 &       2.18 $\pm$       0.16 &       0.10 $\pm$ 
     0.006 &      62.11 $\pm$       8.28 &           73 &       0.27 & 
           0 \\
\hline
B20218+35
 &       5.23
 $\pm$       0.03 &      -57.9 $\pm$        0.0 &      14.08 $\pm$       0.10 & 
    0.0112 &       471. &          217 &       1.35 &        0.0 \\
B20218+35
 &       2.25
 $\pm$       0.03 &      -10.4 $\pm$        0.2 &      45.22 $\pm$       0.53 & 
    0.0026 &       881. &         4332 &       1.54 &        0.0 \\
B20218+35
 &       3.89
 &       -4.5 $\pm$        0.2 &       2.77 $\pm$       0.50 &       0.03 $\pm$ 
     0.011 &     116.48 $\pm$       8.88 &          149 &       0.19 & 
           0 \\
B20218+35
 &      12.27
 $\pm$       0.36 &       -0.3 $\pm$        0.0 &      10.40 $\pm$       0.12 & 
    0.0028 &      4331. &        44682 &       2.48 &        0.0 \\
B20218+35
 &       6.84
 &       -0.3 $\pm$        0.9 &       6.24 $\pm$       2.03 &       0.04 $\pm$ 
     0.030 &      49.87 $\pm$      14.82 &          465 &       0.26 & 
           1 \\
B20218+35
 &       8.44
 &        0.7 $\pm$        0.1 &       2.82 $\pm$       0.34 &       0.11 $\pm$ 
     0.031 &      62.21 $\pm$       8.28 &           73 &       0.38 & 
           2 \\
\hline
B20326+27
 &       1.33
 $\pm$       0.03 &      -12.7 $\pm$        0.6 &      64.39 $\pm$       1.51 & 
    0.0037 &       364. &          217 &       0.96 &        1.0 \\
B20326+27
 &      19.44
 &        0.4 $\pm$        0.0 &       4.67 $\pm$       0.07 &       0.30 $\pm$ 
     0.003 &      78.55 $\pm$       8.88 &          149 &       2.12 & 
           0 \\
B20326+27
 &      19.93
 $\pm$       0.14 &        3.0 $\pm$        0.0 &      13.20 $\pm$       0.05 & 
    0.0184 &      1093. &          465 &       5.12 &        0.5 \\
B20326+27
 &      23.00
 &        6.7 $\pm$        0.0 &       4.56 $\pm$       0.06 &       0.34 $\pm$ 
     0.003 &      83.42 $\pm$       8.28 &           73 &       2.49 & 
           1 \\
\hline
B20400+25
 &       1.97
 $\pm$       0.10 &      -48.0 $\pm$        0.1 &       5.04 $\pm$       0.32 & 
    0.0017 &      1189. &        35507 &       0.18 &        0.0 \\
B20400+25
 &       1.91
 $\pm$       0.04 &      -27.4 $\pm$        0.5 &      40.31 $\pm$       1.29 & 
    0.0112 &       171. &          217 &       1.38 &        1.0 \\
B20400+25
 &      17.65
 &        5.0 $\pm$        0.4 &      14.63 $\pm$       0.63 &       0.03 $\pm$ 
     0.002 &     607.24 $\pm$       8.28 &           73 &       4.58 & 
           0 \\
B20400+25
 &      33.18
 &        7.2 $\pm$        0.0 &       2.17 $\pm$       0.04 &       0.31 $\pm$ 
     0.007 &      67.41 $\pm$      14.82 &          465 &       0.88 & 
           2 \\
B20400+25
 &      23.82
 &        8.6 $\pm$        0.1 &       4.79 $\pm$       0.10 &       0.16 $\pm$ 
     0.006 &      54.35 $\pm$       8.88 &          149 &       0.80 & 
           1 \\
B20400+25
 &       1.38
 $\pm$       0.06 &       19.9 $\pm$        0.3 &      13.70 $\pm$       0.88 & 
    0.0033 &       422. &          555 &       0.30 &        0.0 \\
\hline
B20411+34
 &       2.34
 $\pm$       0.27 &      -28.4 $\pm$        4.6 &      48.16 $\pm$       5.23 & 
    0.0136 &       174. &        50696 &       1.90 &        0.0 \\
B20411+34
 &       1.44
 &      -14.4 $\pm$        0.3 &       3.86 $\pm$       0.66 &       0.02 $\pm$ 
     0.004 &      14.83 $\pm$       8.88 &          325 &       0.03 & 
           3 \\
B20411+34
 &      11.85
 $\pm$       0.59 &       -4.8 $\pm$        0.2 &      27.39 $\pm$       0.65 & 
    0.0133 &       899. &        16401 &       6.23 &        0.0 \\
B20411+34
 &       3.88
 &       -3.3 $\pm$        0.2 &       5.27 $\pm$       0.35 &       0.07 $\pm$ 
     0.003 &     106.09 $\pm$       8.28 &          607 &       0.74 & 
           2 \\
B20411+34
 &      18.58
 $\pm$       0.70 &        4.1 $\pm$        0.2 &       8.50 $\pm$       0.26 & 
    0.0080 &      2322. &         1577 &       3.07 &        0.0 \\
B20411+34
 &      35.65
 &        5.1 $\pm$        0.1 &       4.73 $\pm$       0.20 &       0.90 $\pm$ 
     0.128 &      39.17 $\pm$      14.82 &          489 &       3.27 & 
           1 \\
B20411+34
 &      25.16
 &        8.5 $\pm$        1.5 &       6.18 $\pm$       1.27 &       0.18 $\pm$ 
     0.073 &      68.77 $\pm$       5.04 &          835 &       1.53 & 
           0 \\
\hline
NV0157+28
 &       2.80
 $\pm$       0.02 &      -13.6 $\pm$        0.1 &      40.62 $\pm$       0.25 & 
    0.0083 &       340. &          446 &       1.91 &        0.0 \\
NV0157+28
 &       2.40
 $\pm$       0.04 &      -10.4 $\pm$        0.0 &       4.52 $\pm$       0.09 & 
    0.0022 &      1098. &          489 &       0.21 &        0.0 \\
NV0157+28
 &      14.89
 &       -2.7 $\pm$        0.1 &       9.76 $\pm$       0.18 &       0.05 $\pm$ 
     0.001 &     302.10 $\pm$       8.88 &          325 &       2.90 & 
           1 \\
NV0157+28
 &      13.69
 &       -0.5 $\pm$        0.0 &       1.96 $\pm$       0.12 &       0.04 $\pm$ 
     0.002 &      25.80 $\pm$       8.28 &          607 &       0.04 & 
           0 \\
NV0157+28
 &       3.80
 $\pm$       0.10 &        2.7 $\pm$        0.2 &      10.56 $\pm$       0.26 & 
    0.0041 &       932. &          835 &       0.77 &        1.0 \\
\hline
NV0232+34
 &       2.69
 $\pm$       0.03 &       -6.5 $\pm$        0.1 &      39.32 $\pm$       0.37 & 
    0.0110 &       245. &        33787 &       1.68 &        0.0 \\
NV0232+34
 &      20.16
 &       -1.9 $\pm$        0.1 &       2.04 $\pm$       0.16 &       0.13 $\pm$ 
     0.010 &      43.04 $\pm$       8.28 &          607 &       0.22 & 
           0 \\
NV0232+34
 &      22.52
 &       -0.1 $\pm$        0.2 &       8.49 $\pm$       0.71 &       0.06 $\pm$ 
     0.014 &     321.07 $\pm$       8.88 &          325 &       3.10 & 
           2 \\
NV0232+34
 &      22.90
 &        0.4 $\pm$        0.0 &       1.82 $\pm$       0.09 &       0.20 $\pm$ 
     0.012 &      37.38 $\pm$      14.82 &          489 &       0.27 & 
           3 \\
NV0232+34
 &      14.61
 &        3.0 $\pm$        0.1 &       2.11 $\pm$       0.19 &       0.09 $\pm$ 
     0.008 &      30.54 $\pm$       5.04 &          835 &       0.11 & 
           1 \\
\hline
\hline
\end{longtable}
\vskip 2 in


\end{document}